 
 
 
\catcode`\X=12\catcode`\@=11
 
\def\n@wcount{\alloc@0\count\countdef\insc@unt}
\def\n@wwrite{\alloc@7\write\chardef\sixt@@n}
\def\n@wread{\alloc@6\read\chardef\sixt@@n}
\def\r@s@t{\relax}\def\v@idline{\par}\def\@mputate#1/{#1}
\def\l@c@l#1X{\firstpart.#1}\def\gl@b@l#1X{#1}\def\t@d@l#1X{{}}
 
\def\crossrefs#1{\ifx\all#1\let\tr@ce=\all\else\def\tr@ce{#1,}\fi
   \n@wwrite\cit@tionsout\openout\cit@tionsout=\jobname.cit 
   \write\cit@tionsout{\tr@ce}\expandafter\setfl@gs\tr@ce,}
\def\setfl@gs#1,{\def\@{#1}\ifx\@\empty\let\next=\relax
   \else\let\next=\setfl@gs\expandafter\xdef
   \csname#1tr@cetrue\endcsname{}\fi\next}
\def\m@ketag#1#2{\expandafter\n@wcount\csname#2tagno\endcsname
     \csname#2tagno\endcsname=0\let\tail=\all\xdef\all{\tail#2,}
   \ifx#1\l@c@l\let\tail=\r@s@t\xdef\r@s@t{\csname#2tagno\endcsname=0\tail}\fi
   \expandafter\gdef\csname#2cite\endcsname##1{\expandafter
     \ifx\csname#2tag##1\endcsname\relax?\else\csname#2tag##1\endcsname\fi
     \expandafter\ifx\csname#2tr@cetrue\endcsname\relax\else
     \write\cit@tionsout{#2tag ##1 cited on page \folio.}\fi}
   \expandafter\gdef\csname#2page\endcsname##1{\expandafter
     \ifx\csname#2page##1\endcsname\relax?\else\csname#2page##1\endcsname\fi
     \expandafter\ifx\csname#2tr@cetrue\endcsname\relax\else
     \write\cit@tionsout{#2tag ##1 cited on page \folio.}\fi}
   \expandafter\gdef\csname#2tag\endcsname##1{\expandafter
      \ifx\csname#2check##1\endcsname\relax
      \expandafter\xdef\csname#2check##1\endcsname{}%
      \else\immediate\write16{Warning: #2tag ##1 used more than once.}\fi
      \multit@g{#1}{#2}##1/X%
      \write\t@gsout{#2tag ##1 assigned number \csname#2tag##1\endcsname\space
      on page \number\count0.}%
   \csname#2tag##1\endcsname}}
\def\multit@g#1#2#3/#4X{\def\t@mp{#4}\ifx\t@mp\empty%
      \global\advance\csname#2tagno\endcsname by 1 
      \expandafter\xdef\csname#2tag#3\endcsname
      {#1\number\csname#2tagno\endcsnameX}%
   \else\expandafter\ifx\csname#2last#3\endcsname\relax
      \expandafter\n@wcount\csname#2last#3\endcsname
      \global\advance\csname#2tagno\endcsname by 1 
      \expandafter\xdef\csname#2tag#3\endcsname
      {#1\number\csname#2tagno\endcsnameX}
      \write\t@gsout{#2tag #3 assigned number \csname#2tag#3\endcsname\space
      on page \number\count0.}\fi
   \global\advance\csname#2last#3\endcsname by 1
   \def\t@mp{\expandafter\xdef\csname#2tag#3/}%
   \expandafter\t@mp\@mputate#4\endcsname
   {\csname#2tag#3\endcsname\lastpart{\csname#2last#3\endcsname}}\fi}
\def\t@gs#1{\def\all{}\m@ketag#1e\m@ketag#1s\m@ketag\t@d@l p
   \m@ketag\gl@b@l r \n@wread\t@gsin
   \openin\t@gsin=\jobname.tgs \re@der \closein\t@gsin
   \n@wwrite\t@gsout\openout\t@gsout=\jobname.tgs }
\outer\def\localtags{\t@gs\l@c@l}
\outer\def\globaltags{\t@gs\gl@b@l}
\outer\def\newlocaltag#1{\m@ketag\l@c@l{#1}}
\outer\def\newglobaltag#1{\m@ketag\gl@b@l{#1}}
 
\newif\ifpr@ 
\def\m@kecs #1tag #2 assigned number #3 on page #4.%
   {\expandafter\gdef\csname#1tag#2\endcsname{#3}
   \expandafter\gdef\csname#1page#2\endcsname{#4}
   \ifpr@\expandafter\xdef\csname#1check#2\endcsname{}\fi}
\def\re@der{\ifeof\t@gsin\let\next=\relax\else
   \read\t@gsin to\t@gline\ifx\t@gline\v@idline\else
   \expandafter\m@kecs \t@gline\fi\let \next=\re@der\fi\next}
\def\pretags#1{\pr@true\pret@gs#1,,}
\def\pret@gs#1,{\def\@{#1}\ifx\@\empty\let\n@xtfile=\relax
   \else\let\n@xtfile=\pret@gs \openin\t@gsin=#1.tgs \message{#1} \re@der 
   \closein\t@gsin\fi \n@xtfile}
 
\newcount\sectno\sectno=0\newcount\subsectno\subsectno=0
\newif\ifultr@local \def\ultralocal{\ultr@localtrue}
\def\firstpart{\number\sectno}
\def\lastpart#1{\ifcase#1 \or a\or b\or c\or d\or e\or f\or g\or h\or 
   i\or k\or l\or m\or n\or o\or p\or q\or r\or s\or t\or u\or v\or w\or 
   x\or y\or z \fi}

\def\resetall{\global\advance\sectno by 1\subsectno=0
   \gdef\firstpart{\number\sectno}\r@s@t}
\def\resetsub{\global\advance\subsectno by 1
   \gdef\firstpart{\number\sectno.\number\subsectno}\r@s@t}
\def\newsection#1\par{\resetall\vskip0pt plus.3\vsize\penalty-250
   \vskip0pt plus-.3\vsize\bigskip\bigskip
   \message{#1}\leftline{\bf#1}\nobreak\bigskip}
\def\subsection#1\par{\ifultr@local\resetsub\fi
   \vskip0pt plus.2\vsize\penalty-250\vskip0pt plus-.2\vsize
   \bigskip\smallskip\message{#1}\leftline{\bf#1}\nobreak\medskip}
 
\def\t@gsoff#1,{\def\@{#1}\ifx\@\empty\let\next=\relax\else\let\next=\t@gsoff
   \def\@@{p}\ifx\@\@@\else
   \expandafter\gdef\csname#1cite\endcsname{\relax}
   \expandafter\gdef\csname#1page\endcsname##1{?}
   \expandafter\gdef\csname#1tag\endcsname{\relax}\fi\fi\next}
\def\verbatimtags{\ifx\all\relax\else\expandafter\t@gsoff\all,\fi}
 
\def\(#1){\edef\dot@g{\ifmmode\ifinner(\hbox{\noexpand\etag{#1}})
   \else\noexpand\eqno(\hbox{\noexpand\etag{#1}})\fi
   \else(\noexpand\ecite{#1})\fi}\dot@g}
 
\newif\ifbr@ck
\def\eat#1{}
\def\[#1]{\br@cktrue[\br@cket#1'X]}
\def\br@cket#1'#2X{\def\temp{#2}\ifx\temp\empty\let\next\eat
   \else\let\next\br@cket\fi
   \ifbr@ck\br@ckfalse\br@ck@t#1,X\else\br@cktrue#1\fi\next#2X}
\def\br@ck@t#1,#2X{\def\temp{#2}\ifx\temp\empty\let\neext\eat
   \else\let\neext\br@ck@t\def\temp{,}\fi
   \def\teemp{#1}\ifx\teemp\empty\else\rcite{#1}\fi\temp\neext#2X}
\def\resetbr@cket{\gdef\[##1]{[\rtag{##1}]}}
\def\references{\resetbr@cket\newsection References\par}
 
\newtoks\symb@ls\newtoks\s@mb@ls\newtoks\p@gelist\n@wcount\ftn@mber
    \ftn@mber=1\newif\ifftn@mbers\ftn@mbersfalse\newif\ifbyp@ge\byp@gefalse
\def\defm@rk{\ifftn@mbers\n@mberm@rk\else\symb@lm@rk\fi}
\def\n@mberm@rk{\xdef\m@rk{{\the\ftn@mber}}%
    \global\advance\ftn@mber by 1 }
\def\rot@te#1{\let\temp=#1\global#1=\expandafter\r@t@te\the\temp,X}
\def\r@t@te#1,#2X{{#2#1}\xdef\m@rk{{#1}}}
\def\b@@st#1{{$^{#1}$}}\def\str@p#1{#1}
\def\symb@lm@rk{\ifbyp@ge\rot@te\p@gelist\ifnum\expandafter\str@p\m@rk=1 
    \s@mb@ls=\symb@ls\fi\write\f@nsout{\number\count0}\fi \rot@te\s@mb@ls}
\def\byp@ge{\byp@getrue\n@wwrite\f@nsin\openin\f@nsin=\jobname.fns 
    \n@wcount\currentp@ge\currentp@ge=0\p@gelist={0}
    \re@dfns\closein\f@nsin\rot@te\p@gelist
    \n@wread\f@nsout\openout\f@nsout=\jobname.fns }
\def\m@kelist#1X#2{{#1,#2}}
\def\re@dfns{\ifeof\f@nsin\let\next=\relax\else\read\f@nsin to \f@nline
    \ifx\f@nline\v@idline\else\let\t@mplist=\p@gelist
    \ifnum\currentp@ge=\f@nline
    \global\p@gelist=\expandafter\m@kelist\the\t@mplistX0
    \else\currentp@ge=\f@nline
    \global\p@gelist=\expandafter\m@kelist\the\t@mplistX1\fi\fi
    \let\next=\re@dfns\fi\next}
\def\symbols#1{\symb@ls={#1}\s@mb@ls=\symb@ls} 
\def\bigsymbol{\textstyle}
\symbols{\bigsymbol\ast,\dagger,\ddagger,\sharp,\flat,\natural,\star}
\def\ftnumbers{\ftn@mberstrue} \def\ftsymbols{\ftn@mbersfalse}
\def\paginal{\byp@ge} \def\resetftnumbers{\ftn@mber=1}
\def\ftnote#1{\defm@rk\expandafter\expandafter\expandafter\footnote
    \expandafter\b@@st\m@rk{#1}}
 
\long\def\jump#1\endjump{}
\def\ssum{\mathop{\lower .1em\hbox{$\textstyle\Sigma$}}\nolimits}

\def\qed{\nobreak\kern 1em \vrule height .5em width .5em depth 0em}
\def\newneq{\hbox{\rlap{\hbox to 1\wd9{\hss$=$\hss}}\raise .1em 
   \hbox to 1\wd9{\hss$\scriptscriptstyle/$\hss}}}
\def\subsetne{\setbox9 = \hbox{$\subset$}\mathrel{\hbox{\rlap
   {\lower .4em \newneq}\raise .13em \hbox{$\subset$}}}}
\def\supsetne{\setbox9 = \hbox{$\subset$}\mathrel{\hbox{\rlap
   {\lower .4em \newneq}\raise .13em \hbox{$\supset$}}}}
 
\def\vbar{\mathchoice{\vrule height6.3ptdepth-.5ptwidth.8pt\kern-.8pt}
   {\vrule height6.3ptdepth-.5ptwidth.8pt\kern-.8pt}
   {\vrule height4.1ptdepth-.35ptwidth.6pt\kern-.6pt}
   {\vrule height3.1ptdepth-.25ptwidth.5pt\kern-.5pt}}
\def\f@dge{\mathchoice{}{}{\mkern.5mu}{\mkern.8mu}}
\def\b@c#1#2{{\rm \mkern#2mu\vbar\mkern-#2mu#1}}
\def\b@b#1{{\rm I\mkern-3.5mu #1}}
\def\b@a#1#2{{\rm #1\mkern-#2mu\f@dge #1}}
\def\bb#1{{\count4=`#1 \advance\count4by-64 \ifcase\count4\or\b@a A{11.5}\or
   \b@b B\or\b@c C{5}\or\b@b D\or\b@b E\or\b@b F \or\b@c G{5}\or\b@b H\or
   \b@b I\or\b@c J{3}\or\b@b K\or\b@b L \or\b@b M\or\b@b N\or\b@c O{5} \or
   \b@b P\or\b@c Q{5}\or\b@b R\or\b@a S{8}\or\b@a T{10.5}\or\b@c U{5}\or
   \b@a V{12}\or\b@a W{16.5}\or\b@a X{11}\or\b@a Y{11.7}\or\b@a Z{7.5}\fi}}
 
\catcode`\X=11 \catcode`\@=12

\input amstex
\documentstyle{amsppt}
\NoRunningHeads
\loadbold
\magnification=\magstep1
\hcorrection{-.1in}
\vcorrection{.05in}
\pagewidth{6.5truein}
\pageheight{8.92truein}

\localtags
\TagsOnRight

                   
\define\sect#1{\firstpart . {#1}}
\define\eqtag#1{\tag\etag{#1}}
\define\eqcite#1{~\thetag{\ecite{#1}}}
\define\retag#1{\no\rtag{#1}}
\define\recite#1{~\cite{\bf {\rcite{#1}}}}
\define\eq#1{~\thetag{\ecite{#1}}}
 
 
\newglobaltag{f}
\define\midfig#1#2#3{\midspace{#1}\caption{{\it Figure \ftag{#2}. }#3.}}
\define\topfig#1#2#3{\topspace{#1}\caption{{\it Figure \ftag{#2}. }#3.}}
\define\capfig#1#2#3{\medpagebreak{\it caption for Figure \ftag{#2}. }
{\sl #3.}\par\medpagebreak}
\define\nofig#1#2#3{\def\discrd{\ftag{#2}}}

\define\ficite#1{\text{Figure~\fcite{#1}}}
 
\newglobaltag{t}
\define\midtbl#1#2#3{{#1}\caption{{\it Table \ttag{#2}. }#3.}}
\define\captbl#1#2#3{\medpagebreak{\it caption for Table \ttag{#2}. }
{\sl #3.}\par\medpagebreak}
\define\notbl#1#2#3{\def\discrd{\ttag{#2}}}

\define\tbcite#1{\text{Table~\tcite{#1}}}
 
\define\today{\ifcase\month\or January\or February\or March\or
April\or May\or June\or July\or August\or September\or October\or
November\or December\fi\space\number\day, \number\year}
 
\def\({\left(}
\def\){\DOTSX\right)}
\def\[{\left[}
\def\]{\DOTSX\right]}

\define\a{\alpha}

\define\r{\rho}
 
\define\C {\Bbb C}
\define\N {\Bbb N}
\define\R {\Bbb R}
\define\Z {\Bbb Z}
 
\define\AAA{{\bold A}}
\define\BBB{\bold B}

\define\M{\bold M}
\redefine\P{\bold P}

\define\PP{\Cal P}

\define\A#1{{\bold A}_{#1}}

\define\abs#1{\left|{#1}\right|}

\define\norm#1{\left\|#1\right\|}

\define\re#1{\!\restriction_{#1}\,}

\define\Ss#1{S_{#1}(2^s)}
\define\tmod#1{{\rm (mod~$#1$)}}

\define\qf{quantum formalism}

\define\qe{quantum equilibrium}
\define\Qe{Quantum equilibrium}
\define\Sc{Schr\"odinger}
\define\qm{quantum mechanics}
\define\Qm{Quantum mechanics}
\define\qt{quantum theory}

\define\wf{wave function}

\define\h{\hbar}
\define\qb{Bohmian mechanics}
\define\Qb{Bohmian mechanics}
\define\BM{Bohmian mechanics}
\define\psisq{{|\psi|}^2}
\define\psiqsq{{|\psi(q)|}^2}
\define\psis#1{{|\psi(#1)|}^2}
\define\psiqs#1{{|\psi(q,#1)|}^2}
\define\pvec#1,#2{(#1_1,\dots,#1_{#2})}
\define\ovec#1,#2{#1_1,\dots,#1_{#2}}
\define\der#1,#2{\dfrac{d#1}{d#2}}
\define\pd#1,#2{\dfrac{\partial#1}{\partial#2}}
\define\sd#1,#2{\dfrac{d^2#1}{{d#2}^2}}
\define\spd#1,#2{\dfrac{\partial^2#1}{{\partial#2}^2}}
\define\vpsi{v^\psi}
\define\vipsi{{\bold v}^\psi_k}
\define\vicpsi{{\bold v}^{c\psi}_k}
\define\bvpsi{{\bold v}^\psi}
\redefine\Im{\operatorname{Im}}
\define\grad{\operatorname{grad}}
\define\Div{\operatorname{div}}
\define\born{\r=\psisq}
\define\qeh{quantum equilibrium hypothesis}
\define\supp{\operatorname{supp}}
\define\Q{\Cal Q}
\define\TT{\Cal T}
\redefine\PP{\Bbb P}
\define\psisig{\psi_\sigma}
\define\psisigo{\psi_{\sigma_0}}
\define\Xsig{X_{\sigma}}
\define\Xsigo{X_{\sigma_0}}
\define\Ysig{Y_{\sigma}}
\define\Ysigo{Y_{\sigma_0}}
\redefine\M{\Cal M}
\define\quote#1{{\par\openup-1\jot\smallskip\vskip1\jot\tenpoint
\narrower\narrower\noindent{#1}\smallskip\openup1\jot}}
\define\te{thermodynamic equilibrium} 

\newcount\rk
\define\rmks{\rk=0}
\define\rmk{\global\advance\rk by 1 \medpagebreak\noindent\the\rk. }

\NoBlackBoxes
\topmatter
\title Quantum Equilibrium and the Origin of Absolute Uncertainty \endtitle 
\author Detlef D\"urr,
\footnotemark"$^{1,2}$"\  
Sheldon Goldstein,
\footnotemark"$^1$"\ 
and Nino Zangh\'i\footnotemark"$^{1,3}$"
\endauthor

\abstract The \qf\ is a ``measurement'' formalism---a
phenomenological formalism describing certain macroscopic regularities. We
argue that it can be regarded, and best be understood, as arising from \qb,
which is what emerges from \Sc's equation for a system of particles when we
merely insist that ``particles'' means particles. While distinctly
non-Newtonian, \qb\ is a fully deterministic theory of particles in motion,
a motion choreographed by the \wf. We find that a Bohmian universe, though
deterministic, evolves in such a manner that an {\it appearance\/} of
randomness emerges, precisely as described by the \qf\ and given, for
example, by ``$\r=\psisq$.'' A crucial ingredient in our analysis of the
origin of this randomness is the notion of the effective \wf\ of a
subsystem, a notion of interest in its own right and of relevance to any
discussion of \qt. When the \qf\ is regarded as arising in this way, the
paradoxes and perplexities so often associated with (nonrelativistic) \qt\
simply evaporate.

\bigskip
\noindent {\bf KEY WORDS:\/} Quantum randomness; quantum uncertainty;
hidden variables; effective \wf; collapse of the \wf; the measurement
problem; Bohm's causal interpretation of \qt; pilot wave; foundations of
\qm.
\endabstract
\dedicatory Dedicated  to the memory of J. S. Bell.\enddedicatory
\thanks {\it Received October 23, 1991.\/}\endthanks
\endtopmatter
\footnotetext"$^1$"{Department of Mathematics, Rutgers University, New
Brunswick, New Jersey 08903.}
\footnotetext"$^2$"{Current address: Fakult\"at f\"ur
Mathematik, Universit\"at M\"unchen, 8000 M\"unchen 2, Germany.}
\footnotetext"$^3$"{Current address: Dipartimento di Fisica,
Universit\`a di Genova, INFN, 16146 Genova, Italy.}


\document
\resetall\heading\sect{Introduction}\endheading

\openup2\jot
\block\openup-2\jot I am, in fact, rather firmly convinced that the essentially statistical
character of contemporary quantum theory is solely to be ascribed to the
fact that this (theory) operates with an incomplete description of physical
systems. (Einstein, in ref.\,50, p. 666)\endblock\indent
What is randomness? probability? certainty? knowledge? These are old and
difficult questions, and we shall not focus on them here.
Nonetheless, we shall obtain sharp, striking conclusions concerning the
relationship between these concepts.

Our primary concern in this paper lies with the status and origin of
randomness in quantum theory. According to the \qf, measurements performed
on a quantum system with definite wave function $\psi$ typically yield
random results. Moreover, even the specification of the \wf\ of the
composite system including the apparatus for performing the measurement
will not generally diminish this randomness. However, the quantum dynamics
governing the evolution of the \wf\ over time, at least when no measurement
is being performed, and given, say, by \Sc's equation, is completely
deterministic. Thus, insofar as the particular physical processes which we
call measurements are governed by the same fundamental physical laws that
govern all other processes, \footnote{And it is difficult to believe that
this is not so; the very notion of measurement itself seems too imprecise
to allow such a distinction within a fundamental theory, even if we were
otherwise somehow attracted by the granting to measurement of an
extraordinary status.} one is naturally led to the hypothesis that the
origin of the randomness in the results of quantum measurements lies in
random initial conditions, in our ignorance of the complete description of
the system of interest---including the apparatus---of which we know only
the wave function.

But according to orthodox quantum theory, and most nonorthodox
interpretations as well, {\it the complete description\/} of a system is
provided by its wave function alone, and there is no property of the system
beyond its wave function (our ignorance of) which might account for the
observed quantum randomness.  Indeed, it used to be widely claimed, on the
authority of von Neumann\recite{vN}, that such properties, the so called
hidden variables, are impossible, that as a matter of mathematics,
averaging over ignorance cannot reproduce statistics compatible with the
predictions of the \qf.  And this claim is even now not uncommon, despite
the fact that a widely discussed counterexample, the \qt\ of David
Bohm~[\rcite{Bohm1},\rcite{Bohm2}], has existed for almost four
decades.\footnote{For an analysis of why von Neumann's and related
``impossibility proofs'' are not nearly so physically relevant as
frequently imagined, see Bell's article\recite{Bellrmp}. (See also the
celebrated article of Bell\recite{Bellepr} for an ``impossibility proof'' which
does have physical significance. See as well\recite{Bertlmann's socks}.)
For a recent, and comprehensive, account of Bohm's ideas see \recite{Bohmnewbook}.}

We shall call this theory, which will be ``derived'' and described in
detail in Section~3, {\bf \qb\/}.  \Qb\ is a new mechanics, a completely
deter\-ministic---but distinctly non-Newtonian---theory of particles in
motion, with the \wf\ itself guiding this motion. (Thus the ``hidden
variables'' for \qb\ are simply the particle positions themselves.)
Moreover, while its {\it formulation\/} does not involve the notion of
quantum observables, as given by self-adjoint operators---so that its
relationship to the \qf\ may at first appear somewhat obscure---it can in
fact be shown that \qb\ not only accounts for quantum
phenomena~[\rcite{Bohm2},\rcite{Bohm3},\rcite{Bohm84}], but also embodies
the
\qf\ itself as the very expression of its empirical import\recite{op
paper}. (The analysis in the present paper establishes agreement between
\qb\ and the \qf\ without addressing the question of how the {\it
detailed\/} \qf\ naturally emerges---how and why specific operators, such
as the energy, momentum, and angular momentum operators, end up playing the
roles they do, as well as why ``observables'' should rather generally be
identified with self-adjoint operators. We shall answer these questions
in\recite{op paper}, in which a general analysis of measurement from a
Bohmian perspective is presented. We emphasize that the present paper is
not at all concerned directly with measurement per se, not even of
positions.)  That this is so is for the most part quite
straightforward, but it does involve a crucial subtlety which, so far as
we know, has never been dealt with in a completely satisfactory manner.

The subtlety to which we refer concerns the origin of the very randomness
so characteristic of quantum phenomena. The predictions of \qb\ concerning
the results of a quantum experiment can easily be seen to be precisely
those of the \qf, {\it provided\/} it is assumed that prior to the
experiment the positions of the particles of the systems involved are
randomly distributed according to Born's statistical law, i.e., according
to the probability distribution given by $\psisq$. And the difficulty upon
which we shall focus here concerns the status---the justification and
significance---of this assumption within \qb: not just why it should be
satisfied, but also, and perhaps more important, what---in a completely
deterministic theory---it could possibly mean!

In Section 2 we provide some background to \qb, describing its relationship to
other approaches to \qm\ and how in fact it emerges from an analysis of these
alternatives. This section, which presents a rather personal perspective on
these matters, will play no role in the detailed analysis of the later
sections and may be skipped on a first reading of this paper.

The crucial concepts in our analysis of \qb\ are those of {\bf effective
\wf\/} (Section 5) and {\bf \qe\/} (Sections 4, 6, 13, and 14). The latter
is a concept analogous to, but quite distinct from, thermodynamic
equilibrium.  In particular, \qe\ provides us with a precise and natural
notion of {\bf typicality\/} (Section~7), a concept which frequently arises
in the analysis of ``large systems'' and of the ``long time behavior'' of
systems of any size.  For a universe governed by \qb\ it is of course true
that, given the initial \wf\ and the initial positions of all particles,
{\it everything\/} is completely determined and nothing whatsoever is
actually random.  Nonetheless, we show that typical initial configurations,
for the universe as a whole, evolve in such a way as to give rise to the
{\it appearance\/} of randomness, with {\bf empirical distributions\/}
(Sections 7 and 10) in agreement with the predictions of the \qf.

From a general perspective, perhaps the most noteworthy consequence of our
analysis concerns {\bf absolute uncertainty\/} (Section 11). In a universe
governed by \qb\ there are sharp, precise, and irreducible limitations on
the possibility of obtaining knowledge, limitations which can in no way be
diminished through technological progress leading to better means of
measurement. 

This absolute uncertainty is in precise agreement with Heisenberg's
uncertainty principle. But while Heisenberg used uncertainty to argue for
the meaninglessness of particle trajectories, we find that, with
\qb, absolute uncertainty arises as a necessity, emerging as a remarkably
clean and simple consequence of the existence of trajectories. Thus quantum
uncertainty, regarded as an experimental fact, is {\it explained\/} by \qb,
rather than {\it explained away\/} as it is in orthodox quantum theory.
\bigpagebreak

Our analysis covers all of nonrelativistic \qm. However, since our concern
here is mainly conceptual, we shall for concreteness and simplicity
consider only particles without spin, and shall ignore indistinguishability
and the exclusion principle. Spin and permutation symmetry arise naturally
in \qb~[\rcite{Bellrmp},\rcite{der paper},\rcite{Bohm1},\rcite{Quantum
Fluct},\rcite{SGJSP}], and an analysis {\it explicitly\/} taking them into
account would differ from the one given here in no essential way.

In fact, our analysis really depends only on rather general qualitative
features of the structure of abstract \qt, not on the details of any
specific \qt---such as nonrelativistic \qm\ or a quantum field theory.  In
particular, the analysis does not require a particle ontology; a field
ontology, for example, would do just as well.

Our analysis is, however, fundamentally nonrelativistic. It may well be the
case that a fully relativistic generalization of the kind of physics
explored here requires new
concepts~[\rcite{Ascona},\rcite{Bellj},\rcite{Bohm75},\rcite{Stapp}]---if
not new mathematical structures. But if one has not first understood the
nonrelativistic case, one could hardly know where to begin for the
relativistic one.

Perhaps this paper should be read in the following spirit: In order to
grasp the essence of Quantum Theory, one must first completely understand
{\it at least one\/} \qt.
\newline\newline

\resetall\heading\sect{Reality and the role of the \wf}\endheading
\block\openup-2\jot \indent For each measurement one is required to ascribe to the
$\psi$-function a characteristic, quite sudden change, which {\it depends
on the measurement result obtained,\/} and so {\it cannot be forseen;\/}
from which alone it is already quite clear that this second kind of change
of the $\psi$-function has nothing whatever in common with its orderly
development {\it between\/} two measurements. The abrupt change by
measurement...is the most interesting point of the entire theory....For
{\it this\/} reason one can {\it not\/} put the $\psi$-function directly in
place of...the physical thing...because in the realism point of view
observation is a natural process like any other and cannot {\it per se\/}
bring about an interruption of the orderly flow of natural events.
(\Sc\recite{cat paper})\endblock\indent
The conventional wisdom that the \wf\ provides a complete description of a
quantum system is certainly an attractive possibility: other things being
equal, monism---the view that there is but one kind of reality---is perhaps
more alluring than pluralism.  But the problem of the origin of quantum
randomness, described at the beginning of Section~1, already suggests that
other things are not, in fact, equal.

Moreover, \wf\ monism suffers from another serious defect, to which the
problem of randomness is closely related: \Sc's evolution tends to produce
spreading over configuration space, so that the wave function $\psi$ of a
macroscopic system will typically evolve to one supported by distinct, and
vastly different, macroscopic configurations, to a grotesque macroscopic
superposition, even if $\psi$ were originally quite prosaic.  This is
precisely what happens during a measurement, over the course of which the
\wf\ describing the measurement process will become a superposition of
components corresponding to the various apparatus readings to which the
\qf\ assigns nonvanishing probability. And the difficulty with this
conception, of a world {\it completely\/} described by such an exotic \wf,
is not even so much that it is extravagantly bizarre, but rather that this
conception---or better our place in it, as well as that of the random
events which the \qf\ is supposed to govern---is exceedingly
obscure.\footnote{What we have just described is often presented more
colorfully as the paradox of \Sc's cat\recite{cat paper}.}

What has just been said supports, not the impossibility of \wf\ monism, but
rather its incompatibility with the \Sc\ evolution. And the allure of \wf\
monism is so strong that most interpretations of \qm\ in fact involve the
abrogation of \Sc's equation. This abrogation is often merely implicit and,
indeed, is often presented as if it were compatible with the quantum
dynamics. This is the case, for example, when the measurement postulates,
regarded as embodying ``collapse of the wave packet,'' are simply combined
with \Sc's equation in the formulation of \qt. The ``measurement problem''
is merely an expression of this inconsistency.

There have been several recent proposals---for example, by Wigner
\recite{Wigner84}, by Leggett\recite{Leggett}, by Stapp\recite{Stapp}, by
Weinberg\recite{Weinberg} and by Penrose\recite{Penrose}---suggesting
explicitly that the quantum evolution is not of universal validity, that
under suitable conditions, encompassing those which prevail during
measurements, the evolution of the \wf\ is not governed by \Sc's equation
(see also\recite{Wignerconsc}). A common suggestion is that the quantum
dynamics should be replaced by some sort of ``nonlinear'' (possibly
nondeterministic) modification, to which, on the microscopic level, it is
but an extremely good approximation. One of the most concrete proposals
along these lines is that of Ghirardi, Rimini, and Weber\recite{GRW}.

The theory of GRW modifies \Sc's equation by the incorporation of a random
``quantum jump,'' to a macroscopically localized \wf. As an explanation of
the origin of quantum randomness it is thus not very illuminating,
accounting, as it does, for the randomness in a rather ad hoc manner,
essentially by fiat. Nonetheless this theory should be commended for its
precision, and for the light it sheds on the relationship between Lorentz
invariance and nonlocality (see\recite{Bellj}).

A related, but more serious, objection to proposals for the modification of
\Sc's equation is the following: The quantum evolution embodies a deep
mathematical beauty, which proclaims ``Do not tamper! Don't degrade my
integrity!'' Thus, in view of the fact that (the relativistic extension of)
\Sc's equation, or, better, the quantum theory, in which it plays so
prominent a role, has been verified to a remarkable---and
unprecedented---degree, these proposals for the modification of the quantum
dynamics appear at best dubious, based as they are on purely conceptual,
philosophical considerations.

But is \wf\ monism really so compelling a conception that we must
struggle to retain it in the face of the formidable difficulties it entails?
Certainly not! In fact, we shall argue that even if there were no such
difficulties, even in the case of ``other things being equal,'' a
strong case can be made for the superiority of pluralism.

According to (pre-quantum-mechanical) scientific precedent, when new
mathematically abstract theoretical entities are introduced into a theory,
the physical significance of these entities, their
very meaning insofar as physics is concerned, arises from their dynamical
role, from the role they play in (governing) the evolution of the more
primitive---more familiar and less abstract---entities or dynamical
variables.  For example, in classical electrodynamics the {\it meaning\/}
of the electromagnetic field derives solely from the Lorentz force
equation, i.e., from the field's role in governing the evolution of the
positions of charged particles, through the specification of the forces,
acting upon these particles, to which the field gives rise; while in
general relativity a similar statement can be made for the
gravitational metric tensor.  That this should be so is rather
obvious: Why would these abstractions be introduced in the first place, if
not for their relevance to the behavior of {\it something else\/}, which
somehow already has physical significance?

Indeed, it should perhaps be thought astonishing that the \wf\ was not also
introduced in this way---insofar as it is a field on configuration space
rather than on physical space, the \wf\ is an abstraction of even higher
order than the electromagnetic field.

But, in fact, it was! The concept of the \wf\ originated in 1924 with de
Broglie\recite{dB}, who---intrigued by Einstein's idea of the
``Gespensterfeld''---proposed that just as electromagnetic waves are somehow
associated with particles, the photons, so should material particles, in
particular electrons, be accompanied by waves. He conceived of these waves
as ``pilot waves,'' somehow governing the motion of the associated
particles in a manner which he only later, in the late 1920's, made
explicit\recite{dB2}. However, under an onslaught of criticism by Pauli, he
soon abandoned his pilot wave theory, only to return to it more than two
decades later, after his ideas had been rediscovered, extended, and vastly
refined by David Bohm~[\rcite{Bohm1},\rcite{Bohm2}].

Moreover, in a paper written shortly after \Sc\ invented wave mechanics,
Born too explored the hypothesis that the \wf\ might be a ``guiding
field'' for the motion of the electron\recite{Born}. As consequences of this
hypothesis, Born was led in this paper both to his statistical
interpretation of the \wf\ and to the creation of scattering theory. Born
did not explicitly specify a guiding law, but he did insist that the \wf\
should somehow determine the motion of the electron only statistically,
that deterministic guiding is impossible. And, like de Broglie, he later
quickly abandoned the guiding field hypothesis, in large measure owing to
the unsympathetic reception of Heisenberg, who insisted that physical
theories be formulated directly in terms of observable quantities, like
spectral lines and intensities, rather than in terms of microscopic
trajectories.

The Copenhagen interpretation of \qm\ can itself be regarded as giving the
\wf\ a role in the behavior of something else, namely of certain macroscopic
objects, called ``measurement instruments,'' during ``quantum
measurements''~[\rcite{Bohmqt},\rcite{LL}]. Indeed, the most modest
attitude one could adopt towards
\qt\ would appear to be that of regarding it as a phenomenological
formalism, roughly analogous to the thermodynamic formalism, for the
description of certain {\it macroscopic\/} regularities.  But it should
nonetheless strike the reader as somewhat odd that the \wf, which appears to
be the fundamental theoretical entity of the fundamental theory of what we
normally regard as microscopic physics, should be assigned a role on the
level of the macroscopic, itself an imprecise notion, and specifically in
terms, even less precise, of measurements, rather than on the microscopic
level.

Be that as it may, the modest position just described is not a stable one:
It raises the question of how this phenomenological formalism arises from
the behavior of the microscopic constituents of the macroscopic objects
with which it is concerned. Indeed, this very question, in the context of
the thermodynamic formalism, led to the development of statistical mechanics
by Boltzmann and Gibbs, and, with some help from Einstein, eventually to
the (almost) universal acceptance of the atomic hypothesis.

Of course, the Copenhagen interpretation is not quite so modest. It goes
further, insisting upon the {\it impossibility\/} of just such an
explanation of the (origin of the) \qf. On behalf of this claim---which is
really quite astounding in that it raises to a universal level the personal
failure of a generation of physicists to find a satisfactory {\it
objective\/} description of microscopic processes---the arguments which
have been presented are not, in view of the rather dramatic conclusions
that they are intended to establish, as compelling as might have been
expected.  Nonetheless, the very acceptance of these arguments by several
generations of physicists should lead us to expect that, if not impossible,
it should at best be extraordinarily difficult to account for the \qf\ in
objective microscopic terms.

Exhortations to the contrary notwithstanding, suppose that we do seek a
microscopic origin for the \qf, and that we do this by trying to find a
role on the microscopic level for the \wf, relating it to the behavior of
something else. How are we to proceed? A modest proposal: First try the
obvious! Then proceed to the less obvious and, as is likely to be
necessary, eventually to the not-the-least-bit-obvious. We shall implement
this proposal here, and shall show that we need nothing but the
obvious!\footnote{Insofar as nonrelativistic quantum mechanics is
concerned.}

What we regard as the obvious choice of primitive ontology---the basic
kinds of entities that are to be the building blocks of everything
else\footnote{Except, of course, the \wf.}---should by now be
clear: Particles, described by their positions in space, changing with
time---some of which, owing to the dynamical laws governing their
evolution, perhaps combine to form the familiar macroscopic objects of
daily experience.

However, the {\it specific\/} role the \wf\ should play in governing the
motion of the particles is perhaps not so clear, but for this, too, we
shall find that there is a rather obvious choice, which when combined with
\Sc's equation becomes \qb. (That an abstraction such as the \wf, for a
many-particle system a field that is not on physical space but on configuration
space, should be a fundamental theoretical entity in such a theory appears
quite natural---as a compact expression of dynamical principles governing
an evolution of {\it configurations\/}.\footnote{However, {\it with
\wf\ monism,\/} without such a role and, indeed, without particle positions
from which to form configurations, how can we make sense of a field on the
{\it space of configurations\/}? We might well ask {\it ``What
configurations?''\/} (And the \wf\ really is on configuration space---it is
in this representation that \qm\ assumes its simplest form!)})
\newline\newline

\resetall\heading\sect{\Qb}\endheading

\block\openup-2\jot ...in physics the only observations we must consider are position
observations, if only the positions of instrument pointers. It is a great
merit of the de Broglie-Bohm picture to force us to consider this fact. If you
make axioms, rather than definitions and theorems, about the `measurement'
of anything else, then you commit redundancy and risk
inconsistency. (Bell\recite{Bellpw}) \endblock\indent
Consider a quantum system of $N$ particles, with masses $\ovec m,N$ and
position coordinates $\ovec \bold q,N$, whose \wf\ $\psi=\psi(\ovec \bold
q,N,t)$ satisfies \Sc's equation
$$
i\h\pd \psi,t=-\sum_{k=1}^N{\frac{\h^2}{2m_k}\boldsymbol\Delta_k\psi}+V\psi,
\eqtag{se}$$
where $\boldsymbol\Delta_k=\bold\nabla_k\cdot\bold\nabla_k=\spd{\ },{\bold
q_k}$ and $V=V\pvec\bold q,N$ is the potential energy of the system.

Suppose that the \wf\ $\psi$ does not provide a complete description of the
system, that the most basic ingredient of the description of the state at a
given time $t$ is provided by the positions $\ovec\bold q,N$ of its
particles at that time, and that the \wf\ governs the {\it evolution\/} of (the
positions of) these particles.  Insofar as first derivatives are simpler
than higher derivatives, the simplest possibility would appear to be that
the \wf\ determine the {\it velocities\/} $\ovec{{\bold v}^\psi},N$ of all
the particles. Here ${\bold v}^\psi_k\equiv{\bold v}^\psi_k\pvec\bold q,N$
is a velocity vector field, on {\it configuration space\/}, for the $k$-th
particle, i.e.,
$$
\der{\bold q_k},{t}={\bold v}^\psi_k\pvec\bold q,N.
\eqtag{gge}$$
Since \eqcite{se} and \eqcite{gge} are first order differential equations,
it would then follow that the state of the system is indeed given by
$\psi$ and $q\equiv\pvec\bold q,N$---the specification of these variables
at any time would determine them at all times.

Since two \wf s of which one is a nonzero constant multiple of the other
should be physically equivalent, we demand that $\vipsi$ be
homogeneous of degree $0$ as a function of $\psi$, 
$$
\vicpsi=\vipsi
\eqtag{d0}$$
for any constant $c\ne0$. 

In order to arrive at a form for $\vipsi$ we shall use symmetry as our main
guide. Consider first a single free particle of mass m, whose \wf\
$\psi(\bold q)$ satisfies the free \Sc\ equation
$$
i\h\pd \psi,t=-\frac{\h^2}{2m}\boldsymbol\Delta\psi.
\eqtag{fse}$$
We wish to choose $\bvpsi$ in such a way that the system of equations given by
\eqcite{fse} and 
$$
\der{\bold q},t=\bvpsi(\bold q)
\eqtag{1pge}$$
is Galilean and time-reversal invariant.\footnote{Note that a first-order
(Aristotelian) Galilean invariant theory of particle motion may {\it
appear\/} to be an oxymoron.} Rotation invariance, with the
requirement that $\bvpsi$ be homogeneous of degree $0$, yields the form
$$
\bvpsi=\a \frac{\nabla\psi}{\psi},
$$
where $\a$ is a constant scalar, as the simplest possibility. 

This form will not in general be real, so that we should perhaps take real
or imaginary parts. Time-reversal is implemented on $\psi$ by the
involution $\psi\to\psi^*$ of complex conjugation, which renders \Sc's
equation time reversal invariant. If the full system, including
\eqcite{1pge}, is also to be time-reversal invariant, we must thus have
that 
$$
{\bold v}^{\psi^*}=-\bvpsi,
\eqtag{tr}$$
which selects the form

$$
\bvpsi=\a\Im\frac{\nabla\psi}{\psi}
\eqtag{g1pvf}$$
with $\a$ real. 

Moreover the constant $\a$ is determined by requiring full Galilean
invariance: Since $\bvpsi$ must transform like a velocity under boosts,
which are implemented on \wf s by $\psi\mapsto e^{i\frac m\h\bold
v_0\cdot\bold q}\psi$, invariance under boosts requires
that $\a=\frac\h{m}$, so that \eqcite{g1pvf} becomes
$$
\bvpsi=\frac\h{m}\Im\frac{\nabla\psi}{\psi}.
\eqtag{1pvf}$$

For a general $N$-particle system, with general potential energy
$V$, we define the velocity vector field by requiring \eqcite{1pvf} for each
particle, i.e., by letting 
$$
\vipsi=\frac\h{m_k}\Im\frac{\nabla_k\psi}{\psi},
\eqtag{vf}$$
so that \eqcite{gge} becomes
$$
\der{\bold
q_k},{t}=\frac\h{m_k}\Im\frac{\nabla_k\psi}{\psi}\pvec\bold q,N.
\eqtag{ge}$$

We've arrived at {\bf \Qb\/}: for our system of $N$ particles the state is
given by
$$
\(q,\psi\)
\eqtag{state}$$
and the evolution by
$$
\aligned
\der{\bold
q_k},{t}&=\frac\h{m_k}\Im\frac{\nabla_k\psi}{\psi}\pvec\bold q,N \\
i\h\pd \psi,t&=-\sum_{k=1}^N{\frac{\h^2}{2m_k}\boldsymbol\Delta_k\psi}+V\psi.
\endaligned
\eqtag{evol}$$

We note that \qb\ is time-reversal invariant, and that it is
Galilean invariant whenever $V$ has this property, e.g., when $V$ is the sum of
a pair interaction of the usual form,
$$
V\pvec\bold q,N=\sum_{i<j}\phi(|\bold q_i-\bold q_j|).
\eqtag{sf}$$
However, our analysis will not depend on the form of $V$.

Note also that \qb\ depends only upon the Riemannian structure
$g=(g_{ij})=(m_i\delta_{ij})$ defined by the masses of the particles: In
terms of this Riemannian structure, the evolution equations \eq{se} and
\eq{ge} of \qb\ become
$$
\aligned
\der q,t&=\h\Im\frac{\grad\psi}{\psi}(q)\\
i\h\pd \psi,t&=-\frac{\h^2}{2}\Delta\psi+V\psi,
\endaligned
\eqtag{Evol}$$
where $q=\pvec \bold q,N$ is the configuration, and $\Delta$ and grad are,
respectively, the Laplace-Beltrami operator and the gradient on the
configuration space equipped with this Riemannian structure.
\bigpagebreak

While \qb\ shares \Sc's equation with the usual \qf, it might appear that
they have little else in common. After all, the former is a theory of
particles in motion, albeit of an apparently highly nonclassical,
non-Newtonian character; while the observational content of the latter
derives from a calculus of noncommuting ``observables,'' usually regarded
as implying radical epistemological innovations. Indeed, if the coefficient
in the first equation of \eq{evol} were other than $\frac{\h}{m_k}$, i.e.,
for general constants $\a_k$, the corresponding theory would have
little else in common with the \qf. But for the particular choice of
$\a_k$, of the coefficient in \eq{evol}, which defines \qb, the \qf\ itself
emerges as a phenomenological consequence of this theory.

What makes the choice  $\a_k=\frac{\h}{m_k}$  special---apart from Galilean
invariance, which plays little or no role in the remainder of this
paper---is that with this value, the probability distribution on
configuration space given by $\psiqsq$ possesses the property of
equivariance, a concept to which we now turn.

Note well that $\psi$ on the right hand side of \eq{gge} or \eq{ge} is a
solution to \Sc's equation \eq{se} and is thus time-dependent,
$\psi=\psi(t)$. It follows that the vector field $\vipsi$, the right hand
side of \eq{ge}, will in general be (explicitly) time-dependent. Therefore,
given a solution $\psi$ to \Sc's equation, we cannot in general expect the
evolution on configuration space defined by \eq{ge} to possess a stationary
probability distribution, an object which very frequently plays an
important role in the analysis of a dynamical system. 

However, the distribution given by $\psiqsq$ plays a role similar to that
of---and for all practical purposes is just as good as---a stationary one:
Under the evolution $\rho(q,t)$ of probability densities, of ensemble
densities, arising from \eq{ge}, given by the continuity equation
$$
\pd\rho,t+\Div(\r\vpsi)=0
\eqtag{ed}$$
with $\vpsi=\pvec\bvpsi,N$ the configuration space velocity arising from $\psi$
and $\Div$ the divergence on configuration space, the density $\r=\psisq$ is
stationary {\it relative to $\psi$\/}, i.e., $\r(t)$ retains its form as a
functional of  $\psi(t)$. In other words, 
$$
\text{\sl if $\r(q,t_0)=\psiqs{t_0}$ at some time $t_0$, then
$\r(q,t)=\psiqs t$ for all t.\/}
\eqtag{equiv}$$
We say that such a distribution is {\bf equivariant\/}.\footnote{More
generally, and more precisely, we say that a functional $\psi\to\mu^\psi$,
from \wf s to finite measures on configuration space, is {\bf
equivariant\/} if the diagram
$$
\CD
\psi    @>>>    \mu^\psi\\
@VU_tVV         @VVF_t^\psi V\\
\psi_t  @>>>    \mu^{\psi_t}
\endCD
$$
is commutative, where $U_t=e^{-\frac i\h tH}$, with Hamiltonian
$H=-\sum_{k=1}^N{\frac{\h^2}{2m_k}\boldsymbol\Delta_k\psi}+V\psi,$ is the solution
map for \Sc's equation and $F_t^\psi$ is the solution map for the natural
evolution on measures which arises from \eq{ge}, with initial
\wf\ $\psi$. ($F_t^\psi(\mu)$ is the measure to which $\mu$
evolves in $t$ units of time when the initial \wf\ is $\psi$.)}

To see that $\psisq$ is, in fact, equivariant observe that 
$$
J^\psi=\psisq\vpsi
\eqtag{cv}$$
where $J^\psi=\pvec{\bold J^\psi},N$ is the quantum probability current,
$$
\bold J_k^\psi=\frac\h{2im_k}(\psi^*\nabla_k\psi-\psi\nabla_k\psi^*);
\eqtag{pc}$$
thus $\r(q,t)=\psiqs t$ satisfies \eq{ed}.

Now consider a quantum measurement, involving an interaction between a
system ``under observation'' and an apparatus which performs the
``observation.'' Let $\psi$ be the \wf\ and $q=(q_{sys},q_{app})$ the
configuration of the composite system of system and apparatus. Suppose that
prior to the measurement, at time $t_i$, $q$ is random, with probability
distribution given by $\r(q,t_i)=\psiqs{t_i}$. When the measurement has
been completed, at time $t_f$, the configuration at this time will, of
course, still be random, as will typically be the outcome of the
measurement, as given by appropriate apparatus variables, for example, by
the orientation of a pointer on a dial or by the pattern of ink marks on
paper. Moreover, by equivariance, the distribution of the configuration q
at time $t_f$ will be given by $\r(q,t_f)=\psiqs{t_f}$, in agreement with
the prediction of the \qf\ for the distribution of $q$ at this time. In
particular, \qb\ and the quantum formalism then agree on the statistics for
the outcome of the measurement.\footnote{This argument appears to leave
open the possibility of disagreement when the outcome of the measurement is
not configurationally grounded, i.e., when the apparatus variables which
express this outcome are not functions of $q_{app}$. However, the reader
should recall Bohr's insistence that the outcome of a measurement be
describable in classical terms, as well as note that results of measurements
must always be at least {\it potentially\/} grounded configurationally, in
the sense that we can arrange that they be recorded in configurational
terms {\it without affecting the result.\/} Otherwise we could hardly
regard the process leading to the original result as a completed
measurement.}
\newline\newline

\resetall\heading\sect{The problem of \qe}\endheading

\block\openup-2\jot Then for instantaneous macroscopic configurations the pilot-wave
theory gives the same distribution as the orthodox theory, insofar as the
latter is unambiguous. However, this question arises: what is the good of
{\it either\/} theory, giving distributions over a hypothetical ensemble
(of worlds!) when we have only one world. (Bell\recite{Bellqmcosm})\endblock\indent
Suppose a system has \wf\ $\psi$. We shall call the probability distribution
on configuration space given by $\born$ the {\bf \qe\/} distribution. And we
shall say that a system is {\bf in \qe} when its coordinates are ``randomly
distributed'' according to the \qe\ distribution.  As we have seen, when a
system and apparatus are in \qe\  the results of ``measurement'' arising from
the interaction between system and apparatus will conform with the
predictions of the \qf\ for such a measurement.

More precisely(!), we say that a system is in \qe\ when the \qe\
distribution is appropriate for its description. It is a major goal of this
paper to explain what exactly this might mean and to show that, indeed,
when understood properly, it is {\it typically\/} the case that systems are
in \qe. In other words, our goal here is to clarify and justify the {\bf \qe\ 
hypothesis\/}:
\bigskip

\sl When a system has wave function $\psi$, the distribution $\r$ of its
coordinates satisfies
$$
\born.
\eqtag{born}$$
\rm\newline
We shall do this in the later sections of this paper. In the rest of this
section we will elaborate on the {\it problem\/} of \qe.

From a dynamical systems perspective, it would appear natural to attempt to
justify \eq{born} using such notions as ``convergence to equilibrium,''
``mixing,'' or ``ergodicity''---suitably generalized. And if it were in
fact necessary to establish such properties for Bohmian mechanics in order
to justify the \qe\ hypothesis, we could not reasonably expect to succeed,
at least not with any degree of rigor. The problem of establishing good
ergodic properties for nontrivial dynamical systems is extremely difficult,
even for highly simplified, less than realistic, models.

It might seem that Bohmian mechanics rather trivially {\it fails\/} to
possess good ergodic properties, if one considers the motion arising from
the standard energy eigenstates of familiar systems. However, quantum
systems attain such simple \wf s only through complex interactions, for
example with an apparatus during a measurement or preparation procedure,
during which time they are not governed by a {\it simple\/} \wf. Thus the
question of the ergodic properties of \qb\ refers to the motion under
generic, more complex, \wf s. 

We shall show, however, that establishing such properties is neither
necessary nor sufficient for our purposes: That it is not necessary follows
from the analysis in the later sections of this paper, and that it would
not be sufficient follows from the discussion to which we now turn.

The reader may wonder why the \qe\ hypothesis should present any difficulty
at all. Why can we not regard it as an additional postulate, on say initial
conditions (in analogy with equilibrium statistical mechanics, where the
Gibbs distribution is often uncritically accepted as axiomatic)? Then, by
equivariance, it will be preserved by the dynamics, so that we obtain the
\qe\ hypothesis for all times. In fact, when all is said and done, we shall
find that this is an adequate description of the situation {\it provided
the \qe\ hypothesis is interpreted in the appropriate way\/}. But for the
\qeh\ as so far formulated, such an account would be grossly inadequate.

Note first that the \qe\ hypothesis relates objects belonging to rather
different conceptual categories: The right hand side of
\eq{born} refers to a dynamical object, which from the perspective of \qb\ 
is of a thoroughly objective character; while the left refers to a
probability distribution---an object whose {\it physical\/} significance
remains mildly obscure and moderately controversial, and which often is
regarded as having a strongly subjective aspect. Thus, some explanation or
justification is called for.

One very serious difficulty with \eq{born} is that it {\it seems\/} to be
demonstrably false in a great many situations. For example, the \wf---of
system and apparatus---after a measurement (arising from \Sc's equation) is
supported by the set of all configurations corresponding to the {\it
possible\/} outcomes of the measurement, while the probability distribution
at this time is supported only by those configurations corresponding to the
{\it actual\/} outcome, e.g., given by a specific pointer position, a main
point of measurement being to obtain the information upon which this
probability distribution is grounded.

This difficulty is closely related to an ambiguity in the domain of
physical applicability of \qb. In order to avoid inconsistency we must
regard \qb\ as describing the entire universe, i.e., our system should
consist of all particles in the universe: The behavior of parts of the
universe, of subsystems of interest, must arise from the behavior of the
whole, evolving according to \qb. It turns out, as we shall show, that
subsystems are themselves, in fact, frequently governed by
\qb. But if we {\it postulate\/} that subsystems must obey \qb, we ``commit
redundancy and risk inconsistency.''

Note also that the very nature of our concerns---the origin and
justification of (local) randomness---forces us to consider the universal
level: Local systems are not (always and are never entirely) isolated.
Recall that cosmological considerations similarly arise in connection with
the problem of the origin of irreversibility (see R. Penrose\recite{enm}).

Thus, strictly speaking, for \qb\ only the universe has a \wf, since the
{\it complete\/} state of an $N$ particle universe at any time is given by
{\it its\/} \wf\ $\psi$ and the configuration $q=\pvec \bold q,N$ of its
particles.  Therefore the right hand side of the \qe\ hypothesis \eq{born}
is also obscure as soon as it refers to a system smaller than the entire
universe---and the systems to which \eq{born} is normally applied are very
small indeed, typically microscopic.

Suppose, as suggested earlier, we consider \eq{born} for the entire
universe. Then the right hand side is clear, but the left is completely
obscure: Focus on \eq{born} for THE INITIAL TIME. What physical
significance can be assigned to a probability distribution on the initial
configurations for the entire universe? What can be the relevance to physics
of such an ensemble of universes? After all, we have at our disposal only
the particular, actual universe of which we are a part. Thus, even if we
could make sense of the right hand side of \eq{born}, and in such a way
that \eq{born} remains a consequence of the \qeh\ at THE INITIAL TIME, we
would still be far from our goal, appearances to the contrary
notwithstanding.

Since the inadequacy of the \qeh\ regarded as describing an ensemble of
universes is a crucial point, we wish to elaborate. For each choice of
initial universal \wf\ $\psi$ and configuration $q$, a ``history''---past,
present, and future---is completely determined. In particular, the results
of all experiments, including quantum measurements, are determined.

Consider an ensemble of universes initially satisfying \eq{born}, and
suppose that it can be shown that for this ensemble the outcome of a
particular experiment is randomly distributed with distribution given by
the \qf. This would tell us only that if we were to repeat the {\it
very same\/} experiment---whatever this might mean---many times, sampling
from our ensemble of universes, we would obtain the desired distribution.
But this is both impossible and devoid of physical significance: While we
{\it can\/} perform many {\it similar\/} experiments, differing, however,
at the very least, by location or time, we cannot perform the very
same experiment more than once.

What we need to know about, if we are to make contact with physics, is {\it
empirical distributions\/}---actual relative frequencies within an ensemble
of actual events---arising from repetitions of similar experiments,
performed at different places or times, within a single sample of the
universe---the one we are in. In other words, what is physically relevant
is not sampling across an ensemble of universes---across (initial)
$q$'s---but sampling across space and time within a single universe,
corresponding to a fixed (initial) $q$ (and $\psi$).

Thus, to demonstrate the compatibility of \qb\ with the predictions of the
\qf, we must show that for at least some choice of initial universal $\psi$
and $q$, the evolution \eq{evol} leads to an apparently random pattern of
events, with empirical distribution given by the \qf. In fact, we show much
more.

We prove that for {\it every\/} initial $\psi$, this agreement with the
predictions of the \qf\ is obtained for {\it typical\/}---i.e., for the
overwhelming majority of---choices of initial $q$. And the sense of
typicality here is with respect to the only mathematically
natural---because equivariant---candidate at hand, namely, \qe. 

Thus, on the universal level, the physical significance of \qe\ is as a
measure of typicality, and the ultimate justification of the \qeh\ is, as
we shall show, in terms of the statistical behavior arising from a typical
initial configuration.

According to the usual understanding of the \qf, when a system has \wf\
$\psi$, \eq{born} is satisfied {\it regardless of whatever additional
information we might have\/}. When we claim to have established agreement
between \qb\ and the predictions of the \qf, we mean to include this
statement among those predictions. We are thus claiming to have established
that in a universe governed by \qb\ it is in principle impossible to know
more about the configuration of any subsystem than what is expressed by
\eq{born}---despite the fact that for \BM\ the actual configuration is an
{\it objective\/} property, beyond the \wf.

This may appear to be an astonishing claim, particularly since it refers to
knowledge, a concept both vague and problematical, in an essential way.
More astonishing still is this: This uncertainty, of an absolute and
precise character, emerges with complete ease, the structure of \qb\ being
such that it allows for the formulation and clean demonstration of
statistical statements of a purely {\it objective\/} character which
nonetheless imply our claims concerning the irreducible limitations on possible
knowledge {\it whatever this ``knowledge'' may precisely mean, and however
we might attempt to obtain this knowledge\/}, provided it is consistent
with \qb. We shall therefore call this limitation on what can be known {\it
absolute uncertainty\/}.
\newline\newline

\resetall\heading\sect{The effective \wf}\endheading

\block\openup-2\jot {\it No one can understand this theory until he is willing to think
of $\psi$ as a real objective field rather than just a `probability
amplitude.' Even though it propagates not in $3$-space but in
$3N$-space.} (Bell\recite{Bellqmcosm})\endblock\indent
We now commence our more detailed analysis of the behavior of an
$N$-particle non-relativistic universe governed by Bohmian mechanics,
focusing in this section on the notion of the effective \wf\ of a subsystem.
We begin with some notation. 

We shall use $\Psi$ as the variable for the universal \wf, reserving $\psi$
for the effective wave function of a subsystem, the definition and
clarification of which is the aim of this section. By $\Psi_t$ we shall
denote the universal \wf\ at time t. We shall use $q=\pvec\bold q,N$ as the
{\it generic\/} configuration space variable, which, to avoid confusion, we
shall usually distinguish from the {\it actual\/} configuration of the
particles, for which we shall usually use capitals.  Thus we shall write
$\Psi=\Psi(q)$ and shall denote the configuration of the universe at time t
by $Q_t$.

We remind the reader that according to \qb\ the state $(Q_t,\Psi_t)$ of the
universe at time t evolves via
$$\aligned
\der{Q_t},t &=v^{\Psi_t}(Q_t)\\
i\h\der{\Psi_t},t
&=-\sum_{k=1}^N{\frac{\h^2}{2m_k}\boldsymbol\Delta_k\Psi_t}+V\Psi_t,\endaligned 
\eqtag{be}$$
where $v^\Psi=\pvec{\bold v}^\Psi,N$ with ${\bold v}^\Psi_k$ defined by
\eq{vf}. 

For any given subsystem of particles we obtain a splitting
$$
q=(x,y),
\eqtag{spl}$$
with $x$ the generic variable for the configuration of the subsystem and
$y$ the generic variable for the configuration of the complementary
subsystem, formed by the particles not in the given subsystem.  We shall
call the given subsystem the {\it $x$-system\/}, and we shall sometimes call
its complement---the {\it $y$-system\/}---the {\it environment\/} of the
$x$-system.\footnote{While we have in mind the situation in which the
$x$-system consists of a set of particles selected by their labels, what we
say would not be (much) affected if the $x$-system consisted, say, of all
particles in a given region. In fact the splitting \eq{spl} could be more
general than one based upon what we would normally regard as a division
into complementary systems of particles; for example, the $x$-system might
include the center of mass of some collection of particles, while the
$y$-system includes the relative coordinates for this collection.}

Of course, for any splitting \eq{spl} we have a splitting
$$
Q=(X,Y)
\eqtag{Spl}$$
for the actual configuration. And for the \wf\ $\Psi$ we may write
$\Psi=\Psi(x,y)$.

Frequently the subsystem of interest naturally decomposes into smaller
subsystems. For example, we may have 
$$
x=(x_{sys},x_{app}),
\eqtag{sysapp}$$
for the composite formed by system and apparatus, or
$$
x=\pvec x,M,
\eqtag{Msys}$$
for the composite formed from $M$ disjoint subsystems. And, of course, any
of the $x_i$ in \eq{Msys} could be of the form
\eq{sysapp}.

Consider now a subsystem with associated splitting \eq{spl}. We wish to
explore the circumstances under which we may reasonably regard this
subsystem as ``itself having a \wf.'' This will serve as motivation for our
definition of the effective \wf\ of this subsystem. To this end, suppose
first that the universal \wf\ factorizes so that
$$
\Psi(x,y)=\psi(x)\Phi(y).
\eqtag{prod}$$
Then we obtain the splitting 
$$
v^\Psi=(v^\psi,v^\Phi),
\eqtag{vspl}$$
and, in particular, we have that 
$$
\der{X},t=v^{\psi}(X)
\eqtag{X}$$
for as long as \eq{prod} is satisfied. Moreover, to the extent that the
interaction between the $x$-system and its environment can be ignored, i.e.,
that the Hamiltonian
$$
H=-\sum_{k=1}^N{\frac{\h^2}{2m_k}\boldsymbol\Delta_k}+V
\eqtag{H}$$
in \eq{be} can be regarded as being of the form
$$
H=H^{(x)}+H^{(y)}
\eqtag{Hsum}$$
where $H^{(x)}$ and $H^{(y)}$ are the contributions to H arising from terms
involving only the particle coordinates of the $x$-system, respectively, the
$y$-system\footnote{The sense of the approximation expressed by \eq{Hsum} is
somewhat delicate. In particular, \eq{Hsum} should not be regarded as a
condition on $H$ (or $V$) so much as a condition on (the supports of) the
factors $\psi$ and $\Phi$ of the \wf\ $\Psi$ whose evolution is governed by
$H$; namely, that these supports be sufficiently well separated so that all
contributions to $V$ involving both particle coordinates in the support of
$\psi$ and particle coordinates in the support of $\Phi$ are so small that
they can be neglected when $H$ is applied to such a $\Psi$.}, the form
\eq{prod} is preserved by the evolution, with $\psi$, in particular,
evolving via
$$
i\h\der{\psi},t=H^{(x)}\psi.
\eqtag{x}$$

It must be emphasized, however, that the factorization \eq{prod} is
extremely unphysical. After all, interactions between system and
environment, which tend to destroy the factorization \eq{prod}, are
commonplace. In particular, they occur whenever a measurement is performed
on the $x$-system. Thus, the universal \wf\ $\Psi$ should now be of an
extremely complex form, involving intricate ``quantum correlations''
between $x$-system and $y$-system, however simple it may have been
originally!

Note, however, that if 
$$
\Psi=\Psi^{(1)}+\Psi^{(2)}
\eqtag{Psum}$$
with the \wf s on the right having (approximately\footnote{in an
appropriate sense, of course. Note in this regard that the simplest metrics
$d$ on the projective space of rays $\{c\Psi\}$ are of the form
$d(\Psi,\Psi')=\|\frac{\nabla\Psi}{\Psi}-\frac{\nabla\Psi'}{\Psi'}\|$,
where ``$\|\ \|$'' is a norm on the space of complex vector fields on
configuration space. Moreover the metric $d$ is preserved by the space-time
symmetries (when ``$\|\ \|$'' is translation and rotation invariant).})
disjoint supports, then (approximately)
$$
v^\Psi(Q)=v^{\Psi^{(i)}}(Q)
\eqtag{vrestr}$$
for $Q$ in the support of $\Psi^{(i)}$. Of course, by mere linearity, if
$\Psi$ is of the form \eq{Psum} at some time $\tau$, it will be of the same
form
$$
\Psi_t=\Psi_t^{(1)}+\Psi_t^{(2)}
\eqtag{Ptsum}$$
for all t, where $\Psi_t^{(i)}$ is the solution agreeing with $\Psi^{(i)}$
at time $\tau$ of the second equation of \eq{be}. Moreover, if the supports
of $\Psi^{(1)}$ and $\Psi^{(2)}$ are ``sufficiently disjoint'' at this
time, we should expect the approximate disjointness of these supports, and
hence the approximate validity of \eq{vrestr}, to persist for a
``substantial'' amount of time.

Finally, we note that according to orthodox quantum measurement
theory~[\rcite{vN},\rcite{Bohmqt},\rcite{Wignermeas},\rcite{Wignerint}],
after a measurement, or preparation, has been performed on a quantum
system, the
\wf\ for the composite formed by system and apparatus is of the form
$$
\sum_\a{\psi_\a\otimes\phi_\a}
\eqtag{msum}$$
with the different $\phi_\a$ supported by the macroscopically distinct
(sets of) configurations corresponding to the various possible outcomes of
the measurement, e.g., given by apparatus pointer positions. Of course, for
\BM\ the terms of \eq{msum} are not all on the same footing: one of
them, and only one, is selected, or more precisely supported, by the
outcome---corresponding, say, to
$\a_0$---which {\it actually\/} occurs. To emphasize this we may write
\eq{msum} in the form
$$
\psi\otimes\phi+\varPsi^\perp
\eqtag{esum}$$
where $\psi=\psi_{\a_0}$, $\phi=\phi_{\a_0}$, and
$\varPsi^\perp=\sum_{\a\neq\a_0}{\psi_\a\otimes\phi_\a}$. 

Motivated by these observations, we say that a subsystem, with associated
splitting \eq{spl}, has {\bf effective \wf\/} $\psi$ (at a given time) if
the universal \wf\ $\Psi=\Psi(x,y)$ and the actual configuration $Q=(X,Y)$
(at that time) satisfy
$$
\Psi(x,y)=\psi(x)\Phi(y)+\Psi^\perp(x,y)
\eqtag{ewfa}$$
with $\Phi$ and $\Psi^\perp$ having macroscopically disjoint $y$-supports, and
$$
Y\in \supp{\Phi}.
\eqtag{ewfb}$$
Here, by the macroscopic disjointness of the $y$-supports of $\Phi$ and
$\Psi^\perp$ we mean not only that their supports are disjoint but that
there is a macroscopic function of $y$---think, say, of the orientation of
a pointer---whose values for $y$ in the support
of $\Phi$ differ by a {\it macroscopic\/} amount from its
values for $y$ in the support of $\Psi^\perp$.

The reader familiar with quantum measurement theory should convince himself
(see \eq{msum} and \eq{esum}) that our definition of effective \wf\
coincides with the usual practice of the \qf\ in ascribing \wf s to systems
{\it whenever the latter does assign a \wf\/}. In particular, whenever a
system has a \wf\ for orthodox \qt, it has an effective \wf\ for
\BM.\footnote{Note that the $x$-system will not have an
effective \wf---even approximately---when, for example, it belongs to a
larger microscopic system whose effective \wf\ does not factorize in the
appropriate way. Note also that the {\it larger\/} the environment of the
$x$-system, the {\it greater\/} is the potential for the existence of an
effective \wf\ for this system, owing in effect to the greater abundance of
``measurement-like'' interactions with a larger environment (see, for
example, Point 20 of the Appendix and the references therein).} However,
there may well be situations in which a system has an effective \wf\
according to \qb, but the standard \qf\ has nothing to say.  (We say ``may
well be'' because the usual \qf\ is too imprecise and too controversial
insofar as these questions---for which ``collapse of the wave packet'' must
in some ill-defined manner be invoked---are concerned to allow for a more
definite statement.) Readers who are not familiar with quantum measurement
theory can---as a consequence of our later analysis---simply replace
whatever vague notion they may have of the \wf\ of a system with the more
precise notion of effective \wf.

Despite the slight vagueness in the definition of effective \wf, arising
from its reference to the imprecise notion of the macroscopic, the
effective \wf, when it exists, is unambiguous. In fact, it is given
by\footnote{We identify \wf s related by a nonzero constant factor.} the
{\bf conditional \wf\/}
$$
\psi(x)=\Psi(x,Y),
\eqtag{wf}$$
which, moreover, is (almost) always defined (assuming continuity,
which, of course, we must). In fact, the main result of this paper,
concerning the {\it statistical\/} properties of subsystems, remains valid
when the notion of effective \wf\ is replaced by the completely precise,
and less restrictive, formulation provided by the conditional \wf\ 
\eq{wf}.\footnote{We therefore need not be be too concerned here by the
fact that our definition is also somewhat unrealistic, in the sense that in
situations where we would in practice say that a system has wave function
$\psi$, the terms on the right hand side of \eq{ewfa} are only
approximately disjoint, or, what amounts to the same thing, the first term
on the right is only approximately of the product from, though to an
enormously good degree of approximation.} 

Note that by virtue of the first equation of \eq{be}, the velocity vector
field for the $x$-system is generated by its conditional \wf.  However, the
conditional \wf\ will not in general evolve (even approximately) according
to \Sc's equation, even when the $x$-system is dynamically decoupled from
its environment.  Thus \eq{wf} by itself lacks the central {\it
dynamical\/} implications, as suggested by the preliminary discussion, of
our definition
\eq{ewfa}, \eq{ewfb}.  And it is of course from these dynamical
implications that the \wf\ of a system derives much of its physical
significance.\footnote{In this regard note the following: Let
$W^Y(x)=V_I(x,Y)$, where $V_I$ is the contribution to $V$ arising from the
terms which represent interactions between the $x$-system and the
$y$-system, i.e., $H=H^{x}+H^{(y)}+V_I$. Suppose that $W^Y$ does not depend
upon $Y$ for $Y$ in the support of $\Phi$, $W^Y=W$ for $Y\in\supp\Phi$.
Then the effective \wf\ $\psi$ satisfies $i\h\der{\psi},t=(H^{(x)}+W)\psi$.
The reader should think, for example, of a gas confined by the walls of a
box, or of a particle moving among obstacles. The interaction of the gas or
the particle with the walls or the obstacles---which after all are part of
the environment---is expressed thru $W$.}

Note well that the notion of effective \wf, or conditional \wf, is made
possible by the existence of the {\it actual\/} configuration $Q=(X,Y)$ as
well as $\Psi$! (In particular, the effective---or conditional---\wf\ is
{\it objective,\/} while a related notion in Everett's Many-Worlds or
Relative State interpretation of quantum theory\recite{MW} is merely {\it
relative.\/}\footnote{For an incisive critique of the Many-Worlds
interpretation, as well as a detailed comparison with \qb, see
Bell~[\rcite{Bellepw},\rcite{Bellqmcosm}].}) Note also that the conditional
\wf\ is the function of $x$ most naturally arising from $\Psi$ and
$Y$.\footnote{For particles with spin our definition \eq{ewfa}, \eq{ewfb}
needs no essential modification.  However, \eq{wf} would have to be
replaced by $\Psi(x,Y)=\psi(x)\otimes\Phi,$ where ``$\otimes$'' here
denotes the tensor product over the spin degrees of freedom. In particular,
for particles with spin, a subsystem need not have even a conditional \wf.}

We emphasize that the effective \wf---as well as the conditional \wf---is,
like any honest to goodness attribute or objective property, a functional
of state description, here a function-valued functional of $\Psi$ and
$Q=(X,Y)$ which depends on $Q$ only through $Y$. We shall sometimes write
$$
\psi=\psi^{Y,\Psi}
\eqtag{wfl}$$
to emphasize this relationship. For the conditional or effective \wf\ at
time $t$ we shall sometimes write
$$
\psi_t=\psi^{Y_t,\Psi_t}\equiv\psi_t^{Y_t},
\eqtag{wflt}$$
suppressing the dependence upon $\Psi$.

Note that though we speak of $\psi$ as a property of the $x$-system, it
depends not upon the coordinates of the $x$-system but only upon the
environment, a distinctly peculiar situation from a classical perspective.
In fact, it is precisely because of this that the effective \wf\ behaves
like a degree of freedom for the $x$-system which is independent of its
configuration $X$.

Consider now a composite $x=\pvec x,M$ of {\it microscopic\/} subsystems,
with $M$ not too large, i.e., not ``macroscopically large.''
Suppose that (simultaneously) each x$_i$-system has effective \wf\
$\psi_i$. Then the $x$-system has effective \wf\
$$
\psi(x)=\psi_1(x_1)\psi_2(x_2)\cdots\psi_M(x_M),
\eqtag{pwf}$$
in agreement with the \qf.\footnote{As far as the \qf\ is concerned, recall
that from a purely operational perspective, whatever procedure
simultaneously prepares each system in the corresponding quantum state is a
preparation of the product state for the composite. Moreover, an analysis
of such a simultaneous preparation in terms of quantum measurement theory
would, of course, lead to the same conclusion. Note also that if the
$x$-system is described by a density matrix whose reduced density matrix for
each x$_i$-system is given by the \wf\ $\psi_i$, then this density matrix
is itself, in fact, given by the corresponding product \wf.} To see this,
note that for each $i$ we have that
$$
\Psi=\psi_i(x_i)\Phi_i(y_i)+\Psi_i^\perp(x_i,y_i)
\eqtag{idecomp}$$
with $\Phi_i$ and $\Psi_i^\perp$ having macroscopically disjoint
$y_i$-supports and hence, because the x$_i$-systems are microscopic, having
disjoint $y$-supports as well.\footnote{It is at this point that the
condition that $M$ not be ``too large''---so large that $x$ can be used to
form a macroscopic variable---becomes relevant. And while the
problematical situation which worries us here may seem far fetched, it is
not as far fetched as it initially might appear to be. It may be that
SQUIDs, superconducting quantum interference devices, can be regarded as
giving rise to a situation just like the one with which we are concerned,
in which lots of microscopic systems have, say, the same effective \wf, but the
composite does not have the corresponding product as effective \wf. See,
however, the comment following the proof of \eq{pwf}.}
Moreover,
$$
Y\in\supp\Phi_1\cap\supp\Phi_2\cap\cdots\cap\supp\Phi_M,
\eqtag{Ysupp}$$
and for all such $Y$ we have 
$$
\Psi(\ovec x,M,Y)=\psi_i(x_i)\Phi_i(\hat x_i,Y)
\eqtag{iprod}$$
for all i, where $\hat x_i=\pvec x,M$ with $x_i$ missing. It follows by
separation of variables, writing
$$
\Psi(x,Y)=\psi_1(x_1)\cdots\psi_M(x_M)\Phi(x,Y)
\eqtag{pprod}$$
and dividing by $\prod_i{\psi_i}$, that for $Y$ satisfying \eq{Ysupp}
$$
\Psi(x,Y)=\psi_1(x_1)\cdots\psi_M(x_M)\Phi(Y)
\eqtag{Myprod}$$
and, indeed, that the $x$-system has an effective \wf, given by the product
\eq{pwf}.

Note that this result would not in general be valid for conditional \wf s.
In fact, the derivation of \eq{pwf}, which is used for the equal-time
analysis of Section 7, is the only place where more than \eq{wf} is
required for our results, and even here only the more precise consequence
\eq{iprod} is needed.  Moreover, our more general, multitime analysis (see
Sections 8--10) does not appeal to \eq{pwf} and requires only \eq{wf}.

We wish to point out that while the qualifications under which we have
established \eq{pwf} are so mild that in practice they exclude almost
nothing, \eq{pwf} is nonetheless valid in much greater generality. In fact,
whenever it is ``known'' that the subsystems have the $\psi_i$ as their
respective effective \wf s---by investigators, by devices, or by any
records or traces whatsoever---{\it insofar as this ``knowledge'' is
grounded in the environment of the composite system,\/} i.e., is reflected
in $y$, \eq{pwf} follows without further qualification.

Nonetheless, in order better to appreciate the significance of the
qualification ``microscopic'' for \eq{pwf}, the reader should consider the
following unrealistic but instructive example: Consider a pair of
macroscopic systems with the composite system having effective \wf\ $\psi(x)=
\psi_L(x_1)\psi_L(x_2)+\psi_R(x_1)\psi_R(x_2)$, where $\psi_L$ is a \wf\
supported by configurations in which a macroscopic coordinate is ``on the
left,'' and similarly for $\psi_R$. Suppose that $X_1$ and $X_2$ are ``on
the left.'' Then each system has effective \wf\ $\psi_L$.

What \wf\ would the \qf\ assign to, say, system $1$ in the previous example?
Though we can imagine many responses, we believe that the best answer is,
perhaps, that while the \qf\ is for all practical purposes unambiguous, we
are concerned here with one of those ``impractical purposes'' for which the
usual \qf\ is not sufficiently precise to allow us to make any definite
statement on its behalf. In this regard, see Bell\recite{AM}.

We shall henceforth often say ``\wf'' instead of ``effective \wf.''
\vfil
\eject

\resetall\heading\sect{The fundamental conditional probability
formula}\endheading 

\block\openup-2\jot\indent The intellectual attractiveness of a mathematical argument,
as well as the considerable mental labor involved in following it, makes
mathematics a powerful tool of intellectual prestidigitation---a glittering
deception in which some are entrapped, and some, alas, entrappers. Thus,
for instance, the delicious ingenuity of the Birkhoff ergodic theorem has
created the general impression that it must play a central role in the
foundations of statistical mechanics.... The Birkhoff theorem does us the
service of establishing its own inability to be more than a questionably
relevant superstructure upon [the] hypothesis [of absolute
continuity]. (Schwartz\recite{Schwartz})\endblock\indent
We are ready to begin the detailed analysis of the \qeh\ \eq{born}. We
shall find that by employing, purely as a mathematical device, the \qe\
distribution on the universal scale, at, say, THE INITIAL TIME, we obtain
the \qeh\ in the sense of empirical distributions for all scales at all
times. The key ingredient in the analysis is an elementary conditional
probability formula.

Let us now denote the initial universal \wf\ by $\Psi_0$ and the initial
universal configuration by $Q$, and for definiteness let us take THE
INITIAL TIME to be $t=0$. For the purposes of our analysis we shall regard
$\Psi_0$ as fixed and $Q$ as random. More precisely, for given fixed
$\Psi_0$ we equip the space $\Q=\{Q\}$ of initial configurations with the
quantum equilibrium probability distribution
$\P(dQ)={\P}^{\Psi_0}(dQ)={|\Psi_0(Q)|}^2dQ.$ $Q_t$ is then a random
variable on the probability space $\{\Q,\P\}$, since it is determined via
\eq{be} by the initial condition given by $Q_0=Q$ and $\Psi_0$. Thus, for
any subsystem, with associated splitting \eq{spl}, $X_t$, $Y_t$, and
$\psi_t$ are also random variables on $\{\Q,\P\}$, where $Q_t=(X_t,Y_t)$ is
the splitting of $Q_t$ arising from \eq{spl}, and $\psi_t$ is the (conditional)
\wf\ of the $x$-system at time $t$ (see equation \eq{wflt}).\footnote{The
reader may wonder why we don't also treat $\Psi_0$ as random. First of all,
we don't have to---we are able to establish our results for {\it every\/}
initial $\Psi_0$, without having to invoke in any way any randomness in
$\Psi_0$.  Moreover, if it had proven necessary to invoke randomness in
$\Psi_0$, the results so obtained would be of dubious physical
significance, since to account for the nonequilibrium character of our
world, the initial \wf\ must be a nonequilibrium, i.e., ``atypical,'' \wf.
See the discussion in Sections 12--14.}

We wish again to emphasize that, taking into account the discussion in
Section 4, we regard the quantum equilibrium distribution $\P$, at least
for the time being, solely as a mathematical device, facilitating the
extraction of {\it empirical\/} statistical regularities from
\qb\footnote{in a manner roughly analagous to the use of ergodicity in
deriving the {\it pointwise\/} behavior of time averages for dynamical
systems.}, and otherwise {\it devoid of physical significance\/}. (However,
as a {\it consequence \/} of our analysis, the reader, if he so wishes, can
safely also regard $\P$ as providing a measure of {\it subjective\/}
probability for the initial configuration $Q$.\footnote{After all, $\P$
could in fact be {\it somebody's\/} subjective probability for $Q$.})

Note that by equivariance the distribution of the random variable $Q_t$ is
given by ${|\Psi_t|}^2$. It thus follows directly from \eq{ewfa}, and even
more directly from \eq{wf}, that for the conditional probability
distribution of the configuration of a subsystem, given the configuration of
its environment, we have the {\it fundamental conditional probability
formula\/}\footnote{$\psi$ is to be understood as normalized whenever we
write $\psisq$.}
$$
\P(X_t\in dx|Y_t)=|\psi_t(x)|^2dx,
\eqtag{cp}$$
where $\psi_t=\psi_t^{Y_t}$ is the (conditional) \wf\ of the subsystem at
time $t$.  In particular, this conditional distribution on the
configuration of a subsystem depends on the configuration of its
environment only through its \wf---an object of quite independent dynamical
significance.  In other words, {\it $X_t$ and $Y_t$ are conditionally
independent given $\psi_t$.\/} The entire {\it empirical\/} statistical
content of \qb\ flows from \eq{cp} with remarkable ease.

We wish to emphasize that  \eq{cp} involves conditioning on the detailed
microscopic configuration of the environment---far more information than
could ever be remotely accessible. Thus \eq{cp} is extremely strong. Note
that it implies in particular that 
$$
\P(X_t\in dx|\psi_t)=|\psi_t(x)|^2dx,
\eqtag{cwf}$$
which involves conditioning on what we would be minimally expected to know
if we were testing Born's statistical law \eq{born}. However, it
would be very peculiar to know {\it only\/} this---to know no more than the
\wf\ of the system of interest. But \eq{cp} suggests---and we shall show,
see Section 11---that whatever additional information we might
have can be of no relevance whatsoever to the possible value of
$X_t$.\footnote{It immediately follows from \eq{cp} that for random
$\Psi_0$ we have that
$$
\P(X_t\in dx|Y_t,\Psi_0)=|\psi_t(x)|^2dx,
$$
where now $\P(dQ,d\Psi_0)={|\Psi_0(Q)|}^2dQ\,\mu(d\Psi_0)$ with $\mu$ any
probability measure whatsoever on initial \wf s. Moreover \eq{cwf} remains
valid.}
\newline\newline

\resetall\heading\sect{Empirical distributions}\endheading 

\block\openup-2\jot ...a single configuration of the world will show statistical
distributions over its different parts. Suppose, for example, this world
contains an actual ensemble of similar experimental set-ups....it follows
from the theory that the `typical' world will approximately realize quantum
mechanical distributions over such approximately independent components.
The role of the hypothetical ensemble is precisely to permit definition of
the word `typical.' (Bell\recite{Bellqmcosm})\endblock\indent
In this section we present the simplest application of \eq{cp}, to the
empirical distribution on configurations arising from a large collection of
subsystems, all of which have the ``same'' wave function at a common time.
This is the situation relevant to an equal-time test of Born's statistical
law.  In practice the subsystems in our collection would be widely
separated, perhaps even in different laboratories.

Consider $M$ subsystems, with configurations $\ovec x,M$, where $x_i$ are
coordinates relative to a frame of reference convenient for the $i$-th
subsystem. Suppose that with respect to these coordinates each subsystem
has at time $t$ the same \wf\ $\psi$, with the composite $x=\pvec x,M$
having the corresponding product
$$
\psi_t(x)=\psi(x_1)\cdots\psi(x_M)
\eqtag{pr}$$
as its \wf\ at that time. Then applying the fundamental conditional
probability formula to the $x$-system, we obtain
$$
\P\bigl(X_t \in dx \bigm| Y_t=Y\bigr)=\psis{x_1}\cdots\psis{x_M}\,dx_1\cdots dx_M,
\eqtag{pcpf}$$
where $Y_t=Y$ is the configuration of the environment at this time.  In
other words, we find that relative to the conditional probability
distribution $\P_t^Y(dQ)\equiv\P(dQ|Y_t=Y)$ given the configuration 
of the environment of the composite system at time $t$, the (actual)
coordinates $\ovec X,M$ of the subsystems at this time form a collection of
independent random variables, identically distributed, with common
distribution $\r_{qe}=|\psi|^2$.

In any test of the \qeh\ \eq{born}, it is the {\bf empirical distribution\/}
$$
\r_{emp}(z)=\frac1M\sum_{i=1}^M \delta(z-X_i)
\eqtag{emp}$$
of $\pvec X,M$ which is {\it directly\/} observed---so that the operational
significance of the \qeh\ is that $\r_{emp}$ be (approximately) given by
$\r_{qe}$. Notice that $\r_{emp}$ is a (distribution-valued) random
variable on $(\Q,\P)$, and that
$\r_{emp}(\Gamma)\equiv\int_{\Gamma}\r_{emp}(z)\,dz$ is the relative
frequency in our ensemble of subsystems of the event ``$X_i\in\Gamma$''.

It now follows from the weak law of large numbers that when the number $M$
of subsystems is large, $\r_{emp}$ is very close to $\r_{qe}$ for
($\P_t^Y$-)most initial configurations $Q\in\Q_t^Y\equiv\bigl\{Q\in\Q
\bigm|Y_t=Y \bigr\}$, the {\it fiber\/} of $\Q$ for which $Y_t=Y$: For any
bounded function $f(z)$, and any $\epsilon>0$, let the {\it ``agreement
set''\/} $\AAA(M,f,\epsilon,t)\subset\Q_t^Y$ be the set of initial
configurations $Q\in\Q_t^Y$ for which
$$
\aligned
\|\r_{emp}-\r_{qe}\|_f&\equiv\left|\int\(\r_{emp}(z)-\r_{qe}(z)\)f(z)\,dz\right|\\
&=\biggl|\frac1M\sum_{i=1}^M{f(X_i)}-\int
f(z)\,\psis z\,dz\biggr|\\&\leq\epsilon.\endaligned  
\eqtag{Af}$$
(We suppress the dependence of $\AAA$ upon Y and on the subsystems
under consideration.) Then by the weak law of large numbers 
$$
\P_t^Y\bigl(\AAA(M,f,\epsilon,t)\bigr)=1-\delta(M,f,\epsilon)
\eqtag{agreef}$$
where $\delta\to0$ as $M\to\infty$.

For a single function $f$, $\|\ \|_f$ cannot provide a very good measure of
closeness. Therefore, consider any finite collection $\bold f=(f_\a)$ of
bounded functions, corresponding for example to a coarse graining of value
space, and let
$$
\aligned
\AAA(M,\bold f,\epsilon,t)&\equiv\bigcap_\a
\AAA(M,f_\a,\epsilon,t)\\&\equiv\left\{Q\in\Q_t^Y\left|\,\|\r_{emp}-\r_{qe}\|_{\bold
f}\equiv\sup_\a\|\r_{emp}-\r_{qe}\|_{f_\a}\leq\epsilon\right.\right\}.
\endaligned
\eqtag{Adef}$$
It follows from \eq{agreef} that
$$
\P_t^Y\bigl(\AAA(M,\bold f,\epsilon,t)\bigr)=1-\delta(M,\bold f,\epsilon)
\eqtag{agreebf}$$
where $\delta(M,\bold f,\epsilon)\leq\sum_\a\delta(M,f_\a,\epsilon).$

The empirical distribution $\r_{emp}$ does not probe in a significant way the
joint distribution \eq{pcpf}, i.e., the independence, of $\ovec X,M$---the
law of large numbers is valid under conditions far more general than
independence. To explore independence one might employ pair functions
$f(X_i,X_j)$, or functions of several variables, in a manner analogous to
that of the preceding analysis. Rather than proceeding in this way, we
merely note---more generally---the following:

For any decision regarding the joint distribution of the $X_i$, we have
at our disposal only the values which happen to occur. On the basis of some
feature of these values, we must arrive at a (possibly rather tentative)
conclusion. With any such feature we may associate a subset $\TT$ of the space
$\R^{DM}=\bigl\{\pvec x,M\bigr\}$ of possible joint values, where
$D=\operatorname{dim}(X_i)$ is the dimension of our subsystems.

Let $\Cal T\subset\Bbb R^{DM}$ be a {\it statistical test\/}
for the hypothesis that $\ovec X,M$ are independent, with distribution
$\psisq$.  This means that the failure to occur of the event $\pvec
X,M\in\Cal T$ can be regarded as a strong indication that $\ovec X,M$ are
not generated by such a joint distribution; in other words, it means that
$$
\Bbb P(\Cal T)=1-\delta(\Cal T)
\eqtag{test}$$
with $\delta\ll1,$ where $\Bbb
P\pvec{dx},M=\psis{x_1}\cdots\psis{x_M}\,dx_1\cdots dx_M$ is the joint
distribution under examination. $1-\delta(\Cal T)$ is a measure of the
reliability of the test $\Cal T$. 

Let 
$$
\AAA(\Cal T,t)=\bigl\{Q\in\Q_t^Y\bigm|X_t\equiv\pvec X,M \in\Cal T \bigr\}
\eqtag{AAdef}$$
Then, trivially,
$$
\P_t^Y\bigl(\AAA(\Cal T,t)\bigr)=1-\delta(\Cal T);
\eqtag{atest}$$
i.e., the $\P_t^Y$-size of the set of initial configurations in $Q_t^Y$ for
which the test is passed matches precisely the reliability of the test. (We
remind the reader that the {\it existence\/} of useful tests, analogous to,
but more general than, the one defined for example by \eq{Af}, is a
consequence of the weak law of large numbers.) In particular, the size of
$M$ required for $\delta$ in \eq{agreebf} to be ``sufficiently'' small is
precisely the size required for the corresponding test
$$
\TT=\left\{\pvec
x,M\in\R^{DM}\left|\,\sup_\a\biggl|\frac1M\sum_{i=1}^M{f_\a(x_i)}-\int f_\a(z)\,\psis z\,dz\biggr|\leq\epsilon\right.\right\}
\eqtag{ftest}$$
to be ``sufficiently'' reliable.\footnote{See Point 12 of the Appendix.}

Equations \eq{agreef},\eq{agreebf}, and \eq{atest} are valid only for $Y$
as described, i.e., when the $x$-system has (conditional) \wf\
$\psi_t\equiv\psi^{Y,\Psi_t}$ of the form \eq{pr}, with which we are
primarily concerned. We remark, however, that for a general $Y$ these
equations remain valid, provided the agreement sets which appear in them
are sensibly defined in terms of the conditional distribution
$\PP_t^Y(dx)=|\psi^{Y,\Psi_t}(x)|^2\,dx$ of $X_t$ given $Y_t=Y$. For
example, we may let
$$
\AAA(Y,t)=\bigl\{Q\in\Q_t^Y\bigm|X_t\in\TT(\PP_t^Y)\bigr\},
\eqtag{AYt}$$
where, for any distribution $\PP$ (on $\R^{DM}$), $\TT=\TT(\PP)$ is a test
for $\PP$, satisfying \eq{test} with $\delta(\TT)\ll1.$

In terms of such {\it conditioned agreement sets\/} $\AAA(Y,t)$, we may
define an {\it unconditioned agreeement set\/} $\AAA(t)$ by requiring that 
$$
\AAA(t)\cap\Q_t^Y=\AAA(Y,t);
\eqtag{AtQ}$$
directly in terms of the tests $\TT$, 
$$
\AAA(t)=\bigl\{Q\in\Q\bigm|X_t\in\TT(\PP_t^{Y_t})\bigr\}.
\eqtag{At}$$
Corresponding to equations \eq{agreef}, \eq{agreebf}, and \eq{atest} we
then have that 
$$
\P(\AAA(t))=1-\delta(t)
\eqtag{agree}$$
where
$$
\delta(t)=\int\delta(Y_t,t)\,d\P\ll1
\eqtag{delta}$$
with $\delta(Y,t)\equiv\delta(\TT(\PP_t^Y))$.

Having said this, we wish to emphasize that equations \eq{agreef},
\eq{agreebf}, and \eq{atest} (for a general $Y$), expressing the
``largeness'' of the conditioned agreement sets, are much stronger and much
more relevant than the equations \eq{agree},\eq{delta} which we have just
obtained: The original equations demand that the {\it disagreement set\/}
$\BBB(t)=\AAA(t)^c\equiv\Q\setminus\AAA(t)$ be ``small,'' not just for {\it
``most''\/} fibers $\Q_t^Y$ corresponding to the possible environments $Y$
at time $t$, but for {\it all\/} such fibers. Insofar as the actual
environment $Y_t$ at time $t$ might be rather special---for example,
because it describes a world containing (human) life---the fact that
``disagreement'' has ``insignificant probability'' for {\it every\/}
environment, regardless of how special, is quite important.\footnote{Note,
in particular, that for any condition $\Cal C$ on environments implying,
among other things, that the \wf\ of the $x$-system at time $t$ is of the
form \eq{pr}, we have the same statement of the ``smallness'' of the
disagreement set with respect to the conditional distribution given
$Y_t\in\Cal C$.} Indeed, it is the crucial element in our analysis of
absolute uncertainty in Section 11.

We may summarize the conclusion at which we have so far arrived with the
assertion that for \qb\ {\it typical\/} initial configurations
lead to empirical statistics at time t which are governed by the \qf\ (see
the last paragraph of Section~3). Typicality is to be here understood in
the sense of \qe: something is true for {\it typical\/} initial
configurations if the set of initial configurations for which it is false
is small in the sense provided by the \qe\ distribution $\P$ (and the
appropriate conditional
\qe\ distributions $\P_t^Y$ arising from $\P$).

We wish to emphasize the role of equivariance in our analysis. Notice that
equations \eq{Adef}, \eq{agreebf} would remain valid---with $\delta$
small---if, for example, $\r_{qe}$ were replaced by ${|\psi|}^4$, {\it
provided\/} the sense $\P$ of typicality were given, not by ${|\Psi|}^4$
(which is not equivariant), but by the density to which ${|\Psi_t|}^4$ would
(backwards) evolve as the time decreases from $t$ to THE INITIAL TIME $0$.
This distribution, this sense of typicality, would presumably be
extravagantly complicated and exceedingly artificial.

More important, it would depend upon the time $t$ under consideration,
while equivariance provides a notion of typicality that works for all $t$.
In fact, because of this time independence of typicality for \qe, we
immediately obtain the typicality of joint agreement for a not-too-large
collection of times $\ovec t,J$
$$
\P\(\bigcup_j\BBB(t_j)\)\ll1,
\eqtag{uj}$$
as well as the typicality of joint agreement at {\it most\/} times of a
collection of any size. We shall not go into this in more detail here
because equivariance in fact yields results far more powerful than these,
covering the empirical distribution for configurations $\ovec X,M$ referring
to times $\ovec t,M$ which may all be different, to which we now turn. We
shall find that in exploring this general situation, further novelties of
the quantum domain emerge.
\newline\newline

\resetall\head\sect{Multitime experiments: the
problem\footnote{\hbox{Sections 8--10 should perhaps be skipped at first
reading.}}}\endhead

In the previous section we analyzed the joint distribution of the
simultaneous configurations $\ovec X,M$ of $M$ (distinct and disjoint)
subsystems, each of which has the same wave function $\psi$. We would now
like to consider the more general, and more realistic, situation in which
$\ovec X,M$ refer to any $M$ subsystems, some or all of which might in fact
be the same, at respective times $\ovec t,M$, which might all be different.
And we would again like to conclude that suitably conditioned, $\ovec X,M$
are independent, each with distribution given by $\psisq$; this would
imply, precisely as in Section 7, the corresponding results about empirical
distributions and tests.

We shall find, however, that this multitime situation requires
considerably more care than we have so far needed; in particular, what we
might think at first glance we would like to be true, in fact turns out to
be in general false!

To begin to appreciate the difficulty, consider configurations $X_1$ and
$X_2$ referring to the {\it same\/} system but at different times
$t_1<t_2,$ and suppose this system has \wf\ $\psi$ at both of these times.
Can we conclude that $X_1$ and $X_2$ are independent? Of course not! For
example, if the system is suitably isolated between the times $t_1$ and
$t_2$, so that its configuration undergoes an autonomous evolution, then
$X_2$ will in fact be a function of $X_1$; in the simplest case, when the
\wf\ $\psi$ is a ground state, we will in fact have that $X_2=X_1$.

What has just been described is not, however, an instance of disagreement
with the \qf, which concerns only the results of {\it observation\/}---and in
the previous example observation would destroy the isolation upon which the
strong correlation between $X_1$ and $X_2$ was based. Moreover, the
particular difficulty just described is easily remedied by taking
``observation'' into account. However, it is perhaps worth noting that for
the equal-time analysis it was not necessary in any way to take observation
directly into account to obtain agreement with the \qf---$\ovec X,M$ had
the distribution given by the \qf\ regardless of whether these variables
were observed.

A much more serious, and subtle, difficulty arises from the fact that the
\wf\ $\psi_t$ of a system at time $t$ is itself a random variable (see
\eq{wflt}), while we wish to consider situations in which our systems each
have the same (non-random) \wf\ $\psi$. In the equal-time case this
consideration led to no difficulty---and was barely noticed---since
$\psi_t$ is {\it nonrandom relative\/} to the environment $Y_t$ upon which
we there conditioned. For the multitime case, however, it is at first
glance by no means clear how we should capture the stipulation that our
systems each have
\wf\ $\psi$.

One possibility would be to treat this stipulation as further conditioning,
i.e., to consider the conditional distribution of $\ovec X,M$ given, among
other things, that the \wf s $\psi_{t_i}$ of our respective systems at the
respective times $\ovec t,M$ satisfy $\psi_{t_i}=\psi$ for all $i$. This
would be a bad idea! The conditioning just described can affect the
distribution of the configurations $\ovec X,M$ in surprising, and
uncontrollable, ways.

For example, suppose that when the result of an observation of $X_1$ is
``favorable,'' the happy experimenter proceeds somehow to prepare the
second system in state $\psi$ at time $t_2$, while if the result is
``unfavorable,'' the depressed experimenter requires some extra time to
recuperate, and prepares the  second system in state $\psi$ at time
$t_2'>t_2.$ In this situation $X_1$ need not be independent of
$\psi_{t_2}$, so that conditioning on $\psi_{t_2}$ may bias the
distribution of $X_1$.

Moreover, we believe that this example is not nearly so artificial as it
may at first appear. In the real world, of which the experimenters and
their equipment are a part, which experiments get performed where and when
can, and typically will, be correlated with the results of previous
experiments, with each other, and with any number of other factors, such
as, for example, the weather, which we would not normally take into
account. Therefore, stochastic conditioning can be a very tricky business
here, yielding conditional distributions of a surprising, and thoroughly
unwanted, character.

What has just been said suggests that our multitime formulation is, while
nonetheless inadequate, also perhaps not as general as we might want. The
times at which our experiments are performed, and indeed the subsystems
upon which they are performed, may themselves be random, and a more general
formulation, like the one we shall give, should take this into account.
However, we wish to emphasize that, as we shall see, the primary value of
such a ``random system'' formulation is not increased generality.  Rather,
it is first of all simply the case that, strictly speaking, the systems
upon which experiments get performed are, in fact, themselves random---not
just the results, or the state of the system, but the time of the
experiment as well as the specific system, the particular collection of
particles, upon which we focus and act. Furthermore, when we properly take
{\it this\/} into account, the difficulty we have been discussing vanishes!
\newline\newline

\resetall\heading\sect{Random systems}\endheading 

Consider a pair $\sigma=(\pi,T)$, where $T\in\R$ (with $T\geq0$ if THE
INITIAL TIME is $0$) and $\pi$ is a splitting
$$
q=(x,y)\equiv(\pi q, \pi^{\perp}q)
\eqtag{pidef}$$
(see Section 5); we identify $\pi$ with the projection
$\Q\equiv\R^{3N}\rightarrow\R^{3m}$ onto the configuration of the
($m$-particle) $x$-system, with the components of $x\equiv\pi q$ ordered,
say, as in $q$. $\pi$ comes together with $\pi^{\perp}$, the complementary
projection, onto the coordinates of the environment (also ordered as in
$q$). Thus we may identify $\pi$ with the subset of $\{1,\dots,N\}$
corresponding to the particles of the $x$-system. $\sigma$ specifies a
subsystem at a given time, for example, the system upon which we experiment
and the time at which the experiment begins.\footnote{If
indistinguishability were taken into account, our identification of $\pi$
would have to be modified accordingly.  We might then associate it, for
example, with a subset of $\R^3.$ (See footnote 15.)}

Now allow both $T$ and $\pi$ to be random, i.e., allow $T$
to be a real-valued, and $\pi$ to be a projection-valued, function on the
space $\Q$ of initial configurations. ($\pi$ may thus be identified with a
random subset of $\{1,\dots,N\}$.) For $\sigma=(\pi,T)$ we write
$$
X_{\sigma}=\pi Q_T
\eqtag{xs}$$
for the configuration of the system and 
$$
Y_{\sigma}=\pi^{\perp}Q_T
\eqtag{ys}$$
for the configuration of its environment.\footnote{More explicitly, when
$\pi$ and $T$ are random, $X_{\sigma}$ is the random variable
$$
X_{\sigma}(Q)=\pi(Q)\(Q_{T(Q)}\)
\eqtag{xse}$$
and similarly for $Y_{\sigma}$.} 

We say that a pair
$$
\sigma=(\pi,T),
\eqtag{rs}$$
consisting of a random projection and a random time as described, is a {\bf random system\/} provided 
$$
\{\sigma=\sigma_0\}\in\Cal F(Y_{\sigma_0})
\eqtag{rsc}$$
for any (nonrandom) $\sigma_0=(\pi_0,t)$.\footnote{The condition \eq{rsc},
which is formally what we need, technically suffers from ``measure-$0$
defects''---since a random time $T$ will typically be a continuous random
variable, the event $\{\sigma=\sigma_0\}$ will typically have measure $0$,
while conditional probabilities, for which \eq{rsc} is formally utilized,
are strictly defined only up to sets of measure $0$.  This defect can be
eliminated by replacing \eq{rsc} by the condition that for any $t$ there
exist a number $\epsilon_0(t)>0$ such that
$$
\{\pi=\pi_0,\,t-\epsilon\leq T \leq t\}\in\Cal F(Y_{(\pi_0,t)})
$$
for all $0<\epsilon<\epsilon_0(t)$, using which our formal analysis becomes
rigorous via standard continuity-density arguments. (Of course, if time
were discrete no such technicalities would arise.)} Here we use the notation
$\Cal A\in\Cal F(W_1,W_2,\dots)$ to convey that $I_{\Cal A}$, the indicator
function of the event $\Cal A\subset\Q$, is a function of
$W_1,W_2,\dots$.\footnote{More precisely, $\Cal F(W_1,W_2,\dots)$ denotes
the sigma-algebra generated by the random variables $W_1,W_2,\dots$.}

We emphasize that for a random system $\sigma$, the configuration $\Xsig$
($\Ysig$) of the system (of its environment) is {\it doubly
random\/}---$\sigma$ is itself random, and for a given value $\sigma_0$ of
$\sigma$, $X_{\sigma_0}$ ($Y_{\sigma_0}$) is, of course, still random.

The condition \eq{rsc} says that the value of a random system, i.e., the
identity of the particular subsystem and time that it happens to specify,
is reflected in its environment. In practice, this value is expressed by
the state of the experimenters, their devices and records, and whatever
other features of the environment form the basis of its {\it selection\/}.
It is for this reason that we usually fail to notice that our systems are
random: relative to ``ourselves,'' which we naturally don't think of as
random, they are completely determined. Notice also that
\eq{rsc} fits nicely with the notion of the \wf\ of a subsystem, as
expressed, e.g., by \eq{wf}.\footnote{While the preceding informal description
may not appear to discriminate between \eq{rsc} and the perhaps
equally natural condition
$$
\sigma\in\Cal F(Y_{\sigma}),
$$
which we may formally write as 
$$
\{\sigma=\sigma_0\}\in\Cal F(Y_{\sigma}),
\eqtag{rsca}$$
a careful reading should convey \eq{rsc}. The conditions \eq{rsc} and
\eq{rsca} are not, in fact, equivalent, nor even comparable. In practice
both are satisfied, the validity of \eq{rsca} deriving mainly from the
existence of ``clocks.'' We have defined the notion of random system using
only \eq{rsc} because this is what turns out to be relevant for our
analysis. (Note also that, trivially, $\sigma\in\Cal F(\Ysig,\sigma)$.)}

We shall write $\psi_{\sigma}$ for the (effective or conditional) \wf\
of the random system $\sigma$---given  $Q\in\Q$, the \wf\ at time
$T(Q)$ of the system defined by $\pi(Q)$. Using the notation of equation
\eq{wflt}, we have that 
$$
\psi_\sigma=\psi_{T,\pi}^{\Ysig},
\eqtag{psisig}$$
where the subscript $\pi$ makes explicit the dependence of $\psi_t^Y$ upon
the splitting $q=(x,y)$. Note that $\psisig$ is a functional of both
$\sigma$ and $\Ysig$.

The crucial ingredient in our multitime analysis is the observation that
the fundamental conditional probability formula \eq{cp} remains valid for
random systems: For any random system $\sigma$\footnote{The conditioning
here on $\sigma$ can of course be removed if $\sigma\in\Cal F(\Ysig)$ or,
more generally, if $\psisig\in\Cal F(\Ysig)$, e.g., if $\psisig=\psi$ is
constant, i.e., nonrandom.}
$$
\P(\Xsig\in dx|\Ysig,\sigma)=|\psisig(x)|^2dx,
\eqtag{rscp}$$
which can in a sense be regarded as the most compact expression of the
entire \qf. To see this note that for any value $\sigma_0=(\pi_0,t)$ of
$\sigma$, we have that on $\{\sigma=\sigma_0\}$
$$
\aligned
\P(\Xsig\in dx|\Ysig,\sigma)&=\P(\Xsig\in dx|\Ysig,\sigma=\sigma_0)\\
&=\P(\Xsigo\in dx|\Ysigo,\sigma=\sigma_0)\\
&=\P(\Xsigo\in dx|\Ysigo)\equiv\P(X_t\in dx|Y_t)\\
&=|\psi_t(x)|^2dx\equiv|\psisigo(x)|^2dx\\
&=|\psisig(x)|^2dx,
\endaligned
\eqtag{pfrscp}$$
where we have used \eq{cp} and \eq{rsc}, as well as the obvious fact that
$\Xsig$, $\Ysig$, and $\psisig$ agree respectively with $\Xsigo (\equiv X_t)$,
$\Ysigo (\equiv Y_t)$, and $\psisigo (\equiv \psi_t)$ on $\{\sigma=\sigma_0\}$.
\footnote{The reader familiar with stochastic processes should note the
similarity between \eq{rsc} and \eq{rscp} on the one hand, and the notions
of stopping time and the strong Markov property from Markov process theory.
Indeed, \eq{cp} can be regarded as a kind of Markov property, in relation
to which \eq{rscp} then becomes a strong Markov property.}
\newline\newline

\resetall\heading\sect{Multitime distributions}\endheading 

\block\openup-2\jot ...every atomic phenomenon is closed in the sense that its
observation is based on registrations obtained by means of suitable
amplification devices with irreversible functioning such as, for example,
permanent marks on the photographic plate...the quantum-mechanical
formalism permits well-defined applications only to such closed
phenomena... (Bohr, ref.\,22, pp. 73 and 90)\endblock\indent
Now consider a sequence $\sigma_i=(\pi_i,T_i)$, $i=1,\dots,M$, of random
systems, ordered so that (with probability $1$)
$$
T_1\leq T_2\leq\cdots\leq T_M.
\eqtag{T}$$
We write $X_i$ for $X_{\sigma_i}$, $Y_i$ for $Y_{\sigma_i}$, and let 
$$
{\Cal F}_i=\Cal F(Y_{\sigma_i},\sigma_i).
\eqtag{Fi}$$

Suppose that for the \wf\ of the $i$-th system we have
$$
\psi_{\sigma_i}=\psi_i
\eqtag{psii}$$
where $\psi_i$ is {\it nonrandom\/}, i.e., (with probability $1$)
the random \wf\ $\psi_{\sigma_i}$ is the specific \wf\ $\psi_i$. This will
be the case if the requirement that the $i$-th system have \wf\ $\psi_i$
forms part of the basis of selection for this system, i.e., for $\sigma_i$---
for example, if the $i$-th experiment, by prior decision, must be
preceded by a successful preparation of the state $\psi_i$.

Finally, suppose that 
$$
X_i\in {\Cal F}_j\quad\text{for all $i<j$,}
\eqtag{xf}$$
i.e., for all $i<j$ $X_i$ is a function of $Y_j$ and $\sigma_j$. This will
hold, for example, if, with probability $1$, each $X_i$ is measured---if the
$i$-th measurement has not been completed, and the result ``recorded,''
prior to time $T_j$, then the $i$-th system, together with the apparatus
which measures it, must still be isolated at time $T_j$, from
$\sigma_j$ as well as from the rest of its environment, remaining so until
the completion of this measurement.

Notice that since $\psi_j$ is nonrandom, it follows from
\eq{xf} and the fundamental conditional probability formula \eq{rscp} that 
$$
\aligned
\P(X_j\in dx_j|\ovec X,{j-1})&=\P(X_j\in dx_j|Y_j,\sigma_j)\\
&=|\psi_j(x_j)|^2dx_j.
\endaligned
\eqtag{cpx}$$
Thus
$$
\aligned
\P\(X_i\in dx_i,\,i\leq j\)&=\P\(X_i\in dx_i,\,i\leq j-1\)\P\(X_j\in dx_j|X_1=x_1,\dots,X_{j-1}=x_{j-1}\)\\ &=\P\(X_i\in
dx_i,\,i\leq j-1\)|\psi_j(x_j)|^2dx_j\\
&=|\psi_1(x_1)|^2\cdots|\psi_j(x_j)|^2dx_1\cdots dx_j,
\endaligned
\eqtag{pxs}$$
and
$$
\text{{\sl $\ovec X,M$ are independent, with each $X_i$ having distribution
given by ${|\psi_i|}^2$.\/}}
\eqtag{mte}$$

As it stands \eq{mte} is mildly useless, since the
probability distribution $\P$ with respect to which it is formulated does
not take into account any ``prior'' information, some of which we might
imagine to be relevant to the outcomes of our sequence of experiments.
Therefore, it is significant that our entire random system
analysis (including \eq{T}, \eq{psii}, and \eq{xf}) can be
relativized to any set $\Cal M\subset\Q$---i.e., we may replace $(\Q,\P)$
by $(\Cal M,\P^{\Cal M})$ where $\P^{\Cal M}(dQ)=\P(dQ|\Cal M)$---without
essential modification, {\it provided\/} the random systems $\sigma$ under
consideration satisfy
$$
\M\in\Cal F(\Ysig,\sigma).
\eqtag{mf}$$
In particular, \eq{mte} is valid even with respect to $\P^{\M}$ provided that
for all $i$
$$
\M\in{\Cal F}_i.
\eqtag{mfi}$$

We might think of $\M$ as reflecting the ``macroscopic state'' at a time
prior to all of our experiments, though one might argue about whether
\eq{mfi} would then be satisfied. Be that as it may, any event $\M$
describing any sort of prior information to which we could conceivably
have access would be expected to satisfy \eq{mfi}, particularly if this
information were recorded.

Now suppose that $\psi_i=\psi$ for all $i$. Then the joint distribution of
$\ovec X,M$ with respect to $\P^{\M}$ is precisely the same as in the equal
time situation of Section 7.\footnote{Notice that equal-time experiments
are covered by our multitime analysis---all the $T_i$ can be
identical---and in this case \eq{xf} is automatically satisfied. However,
for our earlier equal-time results it was necessary that $\psi$ be the
effective \wf, while here conditional is sufficient.} Since the
analysis there depended only upon this joint distribution, we may draw the
same conclusions concerning empirical distributions and tests as before. We
thus find for our sequence of experiments that {\it typical\/} initial
configurations---typical with respect to $\P$ or $\P^{\M}$---yield
empirical statistics governed by the quantum formalism.

Perhaps this claimed agreement with the \qf\ requires elaboration. We have
been explicitly concerned here only with the statistics governing the
outcomes of {\it position\/} measurements. Now we were also concerned
only with configurations in our equal-time analysis of Section 7. But our
results there directly implied agreement with the \qf\ for the results of
measurements of any observable: 

Our statistical conclusions there were valid regardless of whether or not
the configura\-tions---the $X_i$---were ``measured.'' Thus, for the equal
time case the joint distribution of any functions $Z_i=f_i(X_i)$ of the
configurations must be inherited from the distribution of the $X_i$
themselves. In particular, by considering subsystems of the form
\eq{sysapp}, where the apparatus ``measures the observable''---i.e.,
self-adjoint operator---$\hat Z_i$, with \wf s
$\hat{\psi}_i=\psi_i\otimes\phi_i$ where $\phi_i$ is the initial(ized) \wf\
of the $i$-th apparatus, letting $Z_i$ be the outcome of this ``measurement
of $\hat Z_i$'' and using what we know about the joint distribution of the
$X_i$, it follows that the $Z_i$ are independent, and, as in the last
paragraph of Section 3, that each $Z_i$ must have the distribution provided
by the \qf, namely, that given by the spectral measure $\r_{\hat
Z_i}^{\psi_i}(dz)$ for $\hat Z_i$ in the state $\psi_i$. (For a detailed
account of how this comes about see~[\rcite{Bohm2},\rcite{Bohm84},\rcite{op
paper}].)

The corresponding result for the multitime case does not, in fact, follow
from \eq{mte}. The latter does require that the configurations be
``measured,'' and a ``measurement of ${\hat Z}_i$''
need not involve, and indeed may be incompatible with, a ``measurement'' of
$X_i$.

But, while it does not follow from the {\it result\/} for the $X_i$, the
corresponding result for ``general measurements'' does, in fact, follow
from the {\it analysis\/} for the $X_i$. We need merely suppose for the
$Z_i$ what we did for the $X_i$, namely, that
$$
Z_i\in {\Cal F}_j\quad\text{for all $i<j$,}
\eqtag{zf}$$
to conclude, for the sequence of outcomes $Z_i$ of ``measurements of
observables'' $\hat Z_i$ in states $\psi_i$, that (with respect to
$\P^{\M}$ for $\M$ satisfying \eq{mf})
$$
\text{{\sl $\ovec Z,M$ are independent, with each $Z_i$ having distribution
given by $\r_{\hat Z_i}^{\psi_i}$,\/}}
\eqtag{zte}$$
from which the usual conclusions concerning empirical distributions and
tests follow immediately.\footnote{That $Z_i=f_i(X_i)$
will in fact {\it be\/} the outcome of what would normally be considered a
measurement of $\hat Z_i$ can be expected only if $\psi_i$ is the effective
\wf\ of the $i$-th system, and not merely the conditional \wf: The
functional form of $Z_i$ is based upon the evolution of a system initially
with effective \wf\ $\psi_i$ interacting with a suitable apparatus but
otherwise isolated. However, the conclusion \eq{zte} for $Z_i=f_i(X_i)$ is
valid even for $\psi_i$ merely the conditional \wf, though in this case
$Z_i$ may have little connection with what is actually observed.}

We emphasize that the assumptions \eq{xf}, \eq{zf}, and \eq{mfi} are
minimal. They demand merely that facts about results and initial
experimental conditions not be ``forgotten.'' Thus they are hardly
assumptions at all, but almost the very conditions essential to enable us,
at the conclusion of our sequence of experiments, to talk in an informed
manner about the experimental conditions and results and compare these with
theory.

Moreover, it is not hard to see that if these conditions are relaxed, the
``predictions'' should not be expected to agree with those of the quantum formalism.\footnote{Note that by selectively ``forgetting''
results we can dramatically alter the statistics of those that we have
not ``forgotten.''} This is a striking illustration of the way in which
\qb\ does not {\it merely\/} agree with the \qf, but, eliminating ambiguities,
illuminates, clarifies, and sharpens it.\footnote{The analysis we have
presented does not allow for the possibility that with nonvanishing
probability $T_i=\infty$, i.e., the conditions for the selection of
$\sigma_i$ are never satisfied. Our results extend to this case provided
that $\pvec X,i$ and $\{T_{i+1}<\infty\}$ are conditionally independent
given $\{T_{i}<\infty\}$ for all $i=1,\dots,M-1$, in which case our results
are valid given $\{T_M<\infty\}$. Note that without the aforementioned
conditional independence our results would not be expected to hold:
Suppose, for example, that if the initial results are ``unfavorable,'' the
depressed experimenter destroys humankind, and systems no longer get
prepared properly. Thus, conditioning on $\{T_M<\infty\}$ yields a
``biased'' sample.  The preceding points to perhaps a different, albeit
rather minor, ambiguity in the \qf, of which \qb\ again forces one to take
note, and in so doing to rectify.}
\newline\newline

\resetall\heading\sect{Absolute uncertainty}\endheading 

That the \qeh\ $\born$ conveys the {\it most detailed\/} knowledge {\it
possible\/} concerning the present configuration of a subsystem (of which
the ``observer'' or ``knower'' is not a part---see Point 23 of the
Appendix), what we have called {\bf absolute uncertainty\/}, is implicit in
the results of Sections $7$ and $10$.\footnote{Note, however, that as far
as knowledge of the past is concerned, it is possible to do a good deal
better than what would be permitted by absolute uncertainty for knowledge
of the present: Having prepared our subsystem in a specific
(not-too-localized) quantum state, with known \wf\ $\psi$, we may proceed
to measure the configuration $X$ of this system, thereby obtaining detailed
knowledge of both its \wf\ and its configuration for some {\it past\/}
time. But note well that the determination of the configuration
may---indeed, as we show, must---lead to an appropriate ``collapse'' of
$\psi$, and hence our knowledge of the (present) configuration will be
compatible with $\born$ for the {\it present\/} \wf. (Note also that for
quantum orthodoxy as well it is sometimes argued that knowledge of the past
need not be constrained by the uncertainty principle.)} The key observation
relevant to this conclusion is this: Whatever we may reasonably mean by
knowledge, information, or certainty---and what precisely these do mean is
not at all an easy question---it simply must be the case that the
experimenters, their measuring devices, their records, and whatever other
factors may form the basis for, or representation of, what could
conceivably be regarded as knowledge of, or information concerning, the
systems under investigation, must be a part of or grounded in the
environment of these systems.

The possession by experimenters of such information must thus be reflected
in {\it correlations\/} between the system properties to which this
information refers and the features of the environment which express or
represent this information. We have shown, however, that {\it given\/} its
\wf\ there can be no correlation between (the configuration of) a system
and (that of) its environment, even if the full microscopic environment
Y---itself grossly more than what we could conceivably have access to---is
taken into account.

Because we consider absolute uncertainty to be a very important conclusion,
with significance extending beyond the conceptual foundations of quantum
theory, we shall elaborate on how our results, for both the equal-time and
the general multitime cases, entail this conclusion. The crucial point is
that the possession of knowledge or information implies the existence of
certain features of the environment, an environmentally based selection
criterion, such that systems selected on the basis of this criterion satisfy
the conditions expressed by this information. (For example, when a
measuring device registers, or the associated computer printout records,
that ``${\left|X\right|}<1$'', it should in fact be more or less the case
that ${\left|X\right|}<1$.)

Suppose that our $M$ systems of Section $7$ have been chosen on the basis
of some features of the environment, say by selection from an ensemble
of $M'$ systems, also of the form considered there. The selection criterion can
be based upon any property of the environment $Y_t=Y$ of the original
(preselection) ensemble. (We allow for a rather arbitrary selection
criterion, though in practice selection would of course be quite
constrained. In particular, a realistic selection criterion should, perhaps, be
the ``same'' for each system; i.e., whether or not the $i$-th system is
selected should depend, for all $i$, upon the same property of $Y$
relative to this system. However, we need here no such constraints.)

Since, with respect to $\P_t^Y$, the configurations of the systems of our
original ensemble were independent, with each having distribution given by
$\psisq$, and since our selection criterion is based solely upon the
environment $Y$ of the original ensemble and in no way directly on the
values of the configurations themselves, it follows that the
configurations $\ovec X,M$ of our selected subsystems have precisely the
same distribution (also relative to $\P_t^Y$) as the original ensemble.
Thus, for typical initial universal configurations, the empirical
distribution of configurations across our selected ensemble will be given
(approximately) by $\psisq$, just as for the original ensemble. It follows
that, whatever else it may be, our selection criterion cannot be based upon
what we could plausibly regard as {\it information\/} concerning system
configurations (more detailed than what is already expressed by $\psisq$).

For the general case, of multitime experiments as described in Section
$10$, the analysis is perhaps even simpler. In fact, for this case there is
really nothing to do, beyond observing that any (environmentally based)
selection criterion, whatever it may be, can be incorporated into the
definition of our random systems, as part of the basis for their selection.
It thus follows from the results of Section $10$ that no such criterion can
be regarded as reflecting any information, beyond $\psisq$, about the
configurations of these systems. Therefore, no devices whatsoever, based on
any present or future technology, will provide us with the corresponding
knowledge. {\sl In a Bohmian universe such knowledge is absolutely
unattainable!\/}\footnote{The reader concerned that we have overlooked the
possibility that information may sometimes be grounded in
non-configurational features of the environment, for example in velocity
patterns, should consider the following (recall as well footnote 12):
\roster
\item Knowledge and information are, in fact, almost always, if not always,
configurationally grounded. Examples are hardly necessary here, but we
mention one---synaptic connections in the brain.
\item Dynamically relevant differences between environments, e.g., velocity
differences, which are not instantaneously correlated with configurational
differences quickly generate them anyway. And we need not be concerned with
differences which are not dynamically relevant!
\item Knowledge and information must be communicable if they are to be of 
any social relevance; their content must be stable under
communication.  But communication typically produces configurational
representations, e.g., pressure patterns in sound waves.
\item In any case, in view of the effective product form \eq{ewfa}, when a
system has an effective \wf, the configuration $Y$ provides an exhaustive
description of the state of its environment (aside from the universal \wf\
$\Psi$---and through it $\Phi$---which for convenience of exposition we are
regarding as given---see also footnotes 27 and 31).
\endroster}

We emphasize that we do not claim that knowledge of the detailed
configuration of a system is impossible, a claim that would be manifestly
false. We maintain only that---as a consequence of the fact that the
configuration $X$ of a system and the configuration $Y$ of its environment
are conditionally independent given its \wf\ $\psi$---{\sl all
such knowledge must be mediated by $\psi$\/}. And we emphasize
that a major reason for the not insignificant length of our argument, as
presented in Sections 6-11, was the necessity to extract from the
aforementioned conditional independence analogous conclusions concerning
empirical correlations.

From our conclusion that when a system has \wf\ $\psi$ we cannot know more
about its configuration $X$ than what is expressed by $\psisq$, it follows
trivially that {\it knowledge\/} that its \wf\ is $\psi$ similarly
constrains our knowledge of the configuration. It also trivially follows
that detailed knowledge of $X$, for example that $X\in I$ for a given set
of values $I$, entails detailed conclusions concerning the \wf, for example
that the (conditional) \wf\ of the system is supported by $I$.\footnote{And
even if the system does not have an effective \wf, we have that any density
matrix describing the system must also be ``supported'' by $I$.}

Finally, in order to further sharpen the character of our absolute
uncertainty, one more point must be made. We have focused here primarily on
the {\it statistical\/} aspect of the \wf\ of a system. But any ``absolute
uncertainty'' based solely upon the fact that knowledge of the
configuration $X$ of a system must be mediated by (knowledge of) some
``object,'' in the sense that the distribution of $X$ can be expressed
simply in terms of that ``object,'' may be sorely lacking in substance if
the ``object'' is {\it merely\/} statistical. In such a case, knowledge of
the ``object'' need amount to nothing more than knowledge that $X$ has the
distribution so expressed.

What lends substance to the ``absolute uncertainty'' in \qb---and justifies
our use of that phrase---is the fact that the relevant ``object,'' the \wf\
$\psi$, plays a dual role: it has, in addition to its
statistical aspect, also a dynamical one, as expressed, e.g., in equations
\eq{X} and \eq{x}. Thus, knowledge of the \wf\ of a system, which sharply
constrains our knowledge of its configuration, is knowledge of something in
its own right, something ``real,'' and not merely knowledge that the
configuration has distribution $\psisq$. 

Moreover, the {\it detailed\/} character of this dynamical aspect is such
that a \wf\ with narrow support quickly spreads, owing to the dispersion in
\Sc's equation, to one with broad support, a change which generates a
similar change in the distribution of the configuration. It follows that
the unavoidable price we must pay for sharp knowledge of the present
configuration of a system is at best hazy knowledge of its future
configuration, i.e., of its ``effective velocity.'' In particular, our
absolute uncertainty embodies absolute unpredictability.  More generally,
the usual uncertainty relations for noncommuting ``observables'' become a
corollary of the \qeh\ $\born$ as soon as the dynamical role of the \wf\ is
taken into account; a detailed analysis can be found
in~[\rcite{Bohm2},\rcite{Bohm84},\rcite{op paper}].
\newline\newline

\resetall\heading\sect{Knowledge and nonequilibrium}\endheading 

The alert reader may be troubled that we have established results about
randomness and uncertainty, results of a flavor often associated with
``chaos'' and ``strong ergodic properties,'' without having to invoke any
of the hard estimates and delicate analysis usually required to establish such
properties.  Indeed, our analysis neither used nor referred to any such
properties. How can this be?

The short answer is \qe, with all that the notion of equilibrium entails
and conveys, an answer upon which we shall elaborate in the next section.
Here we would like merely to observe that what is truly remarkable is not
absolute uncertainty, irreducible limitations on what we {\it can\/} know, but
rather that it is possible to know anything at all! 

We take (the possibility of) knowledge, our information gathering and
storing abilities, too much for granted. (And we conclude all too readily
that the unknowable is unreal.) Of course, it is not at all surprising that
we should do so, in view of the essential role such abilities play in our
existence and survival. But that there should arise stable systems
embodying (what can reasonably be regarded as) such abilities is a perhaps
astonishing fact about the way our universe works, about the laws of
nature!

The point is that we, the knowers, are separate and distinct from the
things about which we know, and know in marvelous detail. How can there be,
between completely disjoint entities, sufficiently strong correlations to
allow for a representation in one of these entities of detailed features of
the other?  Indeed, such correlations are absent in thermodynamic
equilibrium. With respect to (any of the distributions describing) global
thermodynamic equilibrium, disjoint systems are more or less independent,
and systems are more or less independent of their environments, facts
incompatible with the existence of knowledge or information.

What renders knowledge at all possible is nonequilibrium. In fact, rather
trivially, the very existence of the devices and records, not to mention
brains, yielding or embodying any sort of information is impossible under
global equilibrium. And, according to Heisenberg, ``every act of
observation is by its very nature an irreversible
process''\recite{Heisenberg}, and thus fundamentally nonequilibrium.

Thus, the very notion of \qe, of equilibrium of configurations relative to
the \wf, already suggests  the unknowability of these
configurations beyond the \wf. Our results merely provide a firm
foundation for this suggestion. What is, however, striking is the
simplicity of the analysis and how absolute and clean are the conclusions. 

Insofar as equilibrium is associated with the impossibility of knowledge,
equilibrium alone does not provide an adequate perspective on  our analysis.
In particular, our results say perhaps
little of physical relevance unless {\it some\/} knowledge is
possible, e.g., of the \wf\ of a particular system, or of the results
of observations. But for this {\it nonequilibrium\/} is essential.
\newline\newline

\resetall\heading\sect{\Qe\ and thermodynamic (non)equilibrium}\endheading

\block\openup-2\jot [In] a complete physical description, the statistical quantum theory
would...take an approximately analogous position to the statistical
mechanics within the framework of classical mechanics. (Einstein, in
ref.\,50, p.672)\endblock\indent
We would like now to place \qe\ within a broader context by comparing it
with classical thermodynamic equilibrium. 

According to the \qeh, when a system has \wf\ $\psi$, the distribution $\r$ of
its configuration is given by
$$
\born.
\eqtag{born14}$$
Similarly, the Gibbs postulate of statistical mechanics asserts that for a
system at temperature $T$, the distribution $\r$ of its phase space point is
given by
$$
\r=\frac{e^{-H/kT}}Z,
\eqtag{Gibbs}$$
where $H$ is the classical Hamiltonian of the system (including, say, the
``wall potential''), $k$ is Boltzmann's constant, and $Z$, the partition
function, is a normalization.

In addition, we found that \eq{born14} assumed sharp mathematical form when
understood as expressing the  conditional probability formula \eq{cp}.
\eq{Gibbs} is perhaps also best regarded as a conditional probability
formula, for the distribution of the phase point of the system given that
of its environment---after all, the Hamiltonian $H$ typically involves
interactions with the environment, and the temperature $T$ (like the
\wf) can be regarded as a function of (the state of)the {\it
environment.\/} (How otherwise would we know the temperature?) Furthermore,
for a rigorous analysis of equilibrium distributions in the thermodynamic
limit---i.e., of (the idealization given by) global thermodynamic
equilibrium---the equations of Dobrushin and
Lanford-Ruelle~[\rcite{D},\rcite{LR}], stipulating that
\eq{Gibbs}---regarded as expressing such a conditional distribution---be
satisfied for all subsystems, often play a defining role.\footnote{However,
for a universe which, like ours, is not in global thermodynamic
equilibrium, there is presumably no probability distribution on initial
phase points with respect to which the probabilities \eq{Gibbs}, for all
subsystems which happen to be ``in thermodynamic equilibrium'' and all
times, are the conditional probabilities given the environments of the
subsystems. In other words, roughly speaking, \eq{Gibbs} is not
equivariant. (See Krylov\recite{Krylov}, as well as the discussion after
\eq{qs}.)}

Moreover, what we have just described is only a part of a deeper and
broader analogy, between the scheme
$$
\text{classical mechanics}\Longrightarrow\text{equilibrium statistical mechanics}\Longrightarrow\text{thermodynamics},
\eqtag{ts}$$
which outlines the (classical) connection between the microscopic level of
description and a phenomenological formalism on the macroscopic level; and
the scheme
$$
\text{\qb}\Longrightarrow\foldedtext{\qe:\newline statistical
mechanics rel-\newline ative to the \wf}\Longrightarrow\text{the \qf},
\eqtag{qs}$$
which outlines the (quantum) connection between the microscopic level and
another phenomenological formalism---the quantum measurement formalism. We
began this section by comparing only the middle components of \eq{ts} and
\eq{qs}, but it is in fact the full schemes which are roughly analogous.

In particular, note that the middle of both schemes concerns the
equilibrium distribution for the complete state description of the
structure on the left with respect to the state for the structure on the
right---the macrostate, as described by temperature (or energy) and, say,
volume; or the quantum state, specified by the \wf. However, the \qf\ does not
live entirely on the macroscopic level, since the \wf\ for, say, an atom is
best regarded as inhabiting (mainly) the microscopic level, at least for
\qb.

The second arrow of \eq{ts} is, of course, associated primarily with the
work of J. Willard Gibbs\recite{Gibbs}; the corresponding arrow of \eq{qs},
upon which we have not focused here, will be the subject of\recite{op
paper}. (See also~[\rcite{Bohm2},\rcite{Bohm84}].) We have here
focused on the first arrow of \eq{qs}, i.e., on deriving the \qeh\ from
\qb. The corresponding arrow of \eq{ts} remains an active area of research,
though it does not appear likely that a comprehensive rigorous analysis
will be forthcoming any time soon.  Conventional wisdom to the contrary
notwithstanding, the problem of the rigorous justification, from first
principles, of the use of the ``standard ensembles,'' i.e., of the
derivation of randomness governed by detailed probabilities, is far more
difficult for classical thermodynamic equilibrium than for quantum theory!

How can this be? How is it possible so easily to derive the \qeh\ from
first principles (i.e., from \qb), while the corresponding result for
thermodynamics---the rigorous derivation of the Gibbs postulate from first
principles---is so very difficult? The answer, we believe, is that ``pure
equilibrium'' is easy, while nonequilibrium, even a little bit, is hard.
In our nonequilibrium universe, systems which happen to be in \te\ are
surrounded by, and arose from, (thermodynamic) nonequilibrium. Thus with
thermodynamic equilibrium we are dealing with {\sl islands of equilibrium
in a sea of nonequilibrium\/}. But with \qe\ we are in effect dealing with
a {\it global\/} equilibrium, albeit relative to the \wf.

What makes nonequilibrium so very difficult is the fact that for nontrivial
dynamics it is extremely hard to get a handle on the evolution of
nonequilibrium ensembles adequate to permit us rigorously to conclude much
of anything concerning the present distribution that would arise from a
given nonequilibrium distribution in the (distant) past.  To establish
``convergence to equilibrium'' for times $t\to\infty$ (mixing) is itself
extremely difficult, but even this would be of little physical relevance,
since we generally deal with, and can survive only during, times much
earlier than the epoch of global \te.

We should perhaps elaborate on why global equilibrium is so easy. A key
aspect of equilibrium is, of course, stationarity---or equivariance. But
how can this be sufficient for our purposes? Mere stationarity is not
normally sufficient in a dynamical system analysis to conclude that typical
behavior embodies randomness governed by the stationary distribution. Such
``almost everywhere''-type assertions usually require the ergodicity of the
dynamics. Why did we not find it necessary to establish some sort of
ergodicity?

The answer, we believe, lies in another critical aspect of the notion of
equilibrium, shared by the schemes \eq{ts} and \eq{qs}, and arising from
the fact that both schemes are concerned with large ``systems,'' with the
thermodynamic limit as it were. In equilibrium, whether quantum or
thermodynamic, most configurations or phase points are ``macroscopically
similar'': quantities given by suitable spatial averages---e.g., density,
energy density, or velocity fluctuations for \te, and empirical
correlations for \qe---are more or less constant over the state space, in
a sense defined by the equilibrium distribution. To say that a system is in
equilibrium is then to say that its configuration or phase point is typical,
in the sense that the values of these spatial averages are typical.

Now while the individual subsystems with which we have been concerned may
be microscopic, our analysis, in fact, is effectively a ``large system
analysis.'' This is manifest in the equal-time analysis of Section 7, and
for the general, multitime analysis it is implicit in our measurability
conditions \eq{xf} and \eq{mf}, which are plausible only for a universe
having a large number of degrees of freedom. Thus, just as for a system
{\it already in \te,\/} we have no need for the ergodicity of the
dynamics---just ``stationarity''---since the kind of behavior we wish to
establish occurs for a huge set of initial configurations, the
``overwhelming majority.''

(It might also be argued that we have, in fact, established for \qb\ a kind
of effective Bernoulliness, and hence an effective ergodicity. And, again,
the fact that we can do this with little work comes from the
``thermodynamic limit'' aspect of our analysis.)

The reader should compare the impossibility of perpetual motion machines,
which is associated with the scheme \eq{ts}, with that of ``knowledge
machines,'' as expressed by absolute uncertainty, associated with the
scheme \eq{qs}. In both cases the existence of devices of a certain
character is precluded by general theoretical considerations---more or less
equilibrium considerations for both---rather than by a detailed analysis
of the workings of the various possible devices.
\newline\newline

\resetall\heading\sect{Global equilibrium beneath nonequilibrium}\endheading

\block\openup-2\jot But to admit things not visible to the gross creatures that we are
is, in my opinion, to show a decent humility, and not just a lamentable
addiction to metaphysics. (Bell\recite{Bellj})\endblock\indent
The schemes \eq{ts} and \eq{qs} refer to different universes, a classical
universe and a quantum (Bohmian) universe. Since our universe happens to be
a quantum one, it would, perhaps, be better to consider, instead of
\eq{ts}, the analogous quantum scheme\footnote{While it can be shown that
in the ``macroscopic limit''
$$
\text{\qb}\Longrightarrow\text{classical mechanics},
$$
a proper understanding of thermodynamics must be in terms of the {\it
actual\/} behavior of the constituents of equilibrium systems, i.e.,
quantum behavior.}
$$
\text{\qb}\Longrightarrow\text{quantum statistical
mechanics}\Longrightarrow\text{thermodynamics.} 
\eqtag{qts}$$
While the second arrow of \eq{qts} is standard, and presumably
nonproblematical, research on the first arrow has not yet reached its infancy.

Note that it would make little sense to ask for a derivation of quantum
statistical mechanics from the first principles provided by {\it
orthodox\/} \qt.  The very meaning of orthodox \qt\ is so entwined with
processes, such as measurements, in which thermodynamic considerations play
a crucial role that it is difficult to imagine where such a derivation
might begin, or, for that matter, what such a derivation could possibly
mean! (And insofar as \qb\ clarifies the meaning and significance of the
\wf\ of a system, and permits a coherent analysis of the microscopic and
macroscopic domains within a common theoretical framework, it may well be
that the last word has not yet been written concerning the connection
represented by the second arrow.)

If nonequilibrium is an essential aspect of our universe, and if
configurations are in \qe, i.e., pure equilibrium relative to the
\wf, what then is the source, in our universe, of nonequilibrium? What is
it that is {\it not in equilibrium\/}? The \wf, of course---both the
universal \wf\ $\Psi$ and, as a consequence, subsystem \wf s $\psi$. At the
same time, the middle of the scheme \eq{qts} can be regarded as concerned
with the distribution of the subsystem \wf\ $\psi$ for subsystems which
happen to be in \te. But by exploiting global thermodynamic {\it
nonequilibrium\/} we are able to see beneath the thermodynamic-macroscopic
level of description, while with global \qe\ there is no quantum
nonequilibrium to reveal the system configuration $X$ beneath the
system \wf\ $\psi$.

It is important, however, not to succumb to the temptation to conclude, as
does Heisenberg\recite{Heisenberg}, that configurations therefore
provide merely an ``ideological superstructure'' best left out of quantum
theory; for, as we have seen, the very meaning of the
\wf\ $\psi$ of a subsystem requires the existence of configurations, i.e.,
those of its environment. And when we determine the \wf\ of a system we do
so on the basis of the configuration of the environment. Recall also that
both aspects of the \wf\ of a subsystem, the statistical and the dynamical,
cannot coherently be formulated without reference to configurations. It is
therefore not at all astonishing that orthodox \qt, by refusing to accept
configurations as part of the description of the state of a system, has
led to so much conceptual confusion.

Note that the fact that thermodynamics seems to depend only upon $\psi$,
and not on any contribution to the total thermodynamic entropy from the
actual configuration $X$, is an immediate consequence of \qe: For a
universe in \qe\ the entropy associated with configurations is maximal,
i.e., constant as a functional of $\psi$, and thus plays no thermodynamic role.

A crucial feature of our quantum universe is the peaceful coexistence
between global equilibrium (quantum) and nonequilibrium (thermodynamic),
providing us with what we may regard as an ``equilibrium laboratory,'' a
glimpse, as it were, of pure equilibrium, with all the surprising
consequences it entails. Our analysis has shown how the interplay between
the corresponding levels of structure---the nonequilibrium level given by
the \wf, and, beneath the level of the \wf, that of the particles,
described by their positions, in equilibrium relative to the \wf---leads to
the randomness and uncertainty so characteristic of \qt. We shall explore
elsewhere\recite{der paper} how this (hierarchical) structure itself
naturally arises, and what its deeper significance might be. (See also
Bohm\recite{Bohm imp order}.)
\bigpagebreak

We have argued, and believe our analysis demonstrates, that quantum
randomness can best be understood as arising from ordinary ``classical''
uncertainty---about what is {\it there\/} but {\it unknown\/}. The denial
of the existence of this unknowable---or only partially knowable---reality
leads to ambiguity, incoherence, confusion, and endless controversy. What
does it gain us?
\newline\newline

\resetall\heading{Appendix: Random points}\endheading 
\def\firstpart{A}

In the following remarks we expand upon concepts introduced in this paper,
placing our conclusions within a broader perspective and comparing ours
with related approaches.

\rmks
\rmk \qb\ is what emerges from \Sc's equation, which is said to describe the
evolution of the \wf\ of a system of {\it particles\/}, when we take this
language seriously, i.e., when we insist that {\sl ``particles'' means
particles\/}. Thus \qb\ is the minimal interpretation of nonrelativistic
\qt, arising as it does from the assertion that a familiar word has its
familiar meaning.

In particular, if \qb\ is somehow strange or unacceptable, it must be
because either \Sc's equation, or the assertion that ``particles'' means
particles, or their combination is strange or unacceptable. Now the
assertion that ``particles'' means particles can hardly be regarded as in
any way problematical. On the other hand, \Sc's equation, for a field on
{\it configuration\/} space, is a genuine innovation, though one that
physicists by now, of course, take quite for granted. However, as we have
seen in Section 2, when it is appropriately combined with the assertion
that ``particles'' means particles, its strangeness is, in fact, very much
diminished.

\rmk \Qm\ is notoriously nonlocal\recite{Sc entanglement}, a novelty which
is in no way ameliorated by \qb. In fact, ``in this theory an explicit
causal mechanism exists whereby the disposition of one piece of apparatus
affects the results obtained with a distant piece''\recite{Bellrmp}. We
wish to emphasize, however, that {\it relative to the \wf\/}, \qb\ is
completely {\it local\/}: the nonlocality in \qb\ derives solely from the
nonlocality built into the structure of standard \qt, as provided by a \wf\
on configuration space.
\block\openup-2\jot That the guiding wave, in the general case, propagates
not in ordinary three-space but in a multidimensional-configuration space
is the origin of the notorious `nonlocality' of \qm. It is a merit of the
de Broglie-Bohm version to bring this out so explicitly that it cannot be
ignored. (Bell\recite{Belldm})\endblock

\rmk A rather fortunate property of \qb\ is that the behavior of the
parts---of subsystems---reflects that of the whole. Indeed, if this were
not the case it would have been difficult, if not impossible, to have ever
discovered the full theory. We believe that a major reason nonlocality is
so often regarded as problematical is not nonlocality per se but rather
that it {\it suggests\/} the breakdown of precisely this feature.

\rmk Notice that the effective \wf\ $\psi$ is, in effect, a ``collapsed''
\wf. Thus our analysis implicitly explains the status and role of
``collapse of the wave packet'' in the \qf. (See also Point 21, recalling
that the Wigner formula\recite{Wignermeas} for the joint distribution of
the outcomes of a sequence of quantum measurements, to which we there
refer, is usually based upon collapse.)

In particular, note that the effective \wf\ of a subsystem evolves according
to \Sc's equation only when this system is suitably isolated. More
generally, the evolution $\psi(t)$ of the effective \wf\ defines a
stochastic process, one which embodies collapse in just the right
way---with respect to the conditional probability distribution given the
(initial) configuration of the environment of the composite system which
includes the apparatus, with $\psi$ the effective \wf\ of the system alone,
i.e., not including the apparatus. For details see\recite{op paper}.

Note also that the very notion of effective \wf, as well as its behavior,
depends upon the location of the split between the ``observed'' and the
``observer,'' i.e., between the system of interest and the rest of the
world, a dependence whose importance has been emphasized by
Bohr\recite{Bohr}, by von Neumann\recite{vN}, and by a great many others,
see for example~[\rcite{Bohmqt},\rcite{LL},\rcite{LB}].  In
particular, while the effective
\wf\ will ``collapse'' during measurement if the apparatus is {\it not\/}
included in the system, it need not, in principle, collapse if the
apparatus {\it is\/} included, precisely as emphasized by von
Neumann\recite{vN}.  But von Neumann was left with the ``measurement
paradox,'' while with \qb\ no hint of paradox remains.

\rmk The fact that knowledge of the configuration of a system must be
mediated by its \wf\ may partially account, from a Bohmian perspective, for
how the physics community could identify the state of a quantum
system---its complete description---with its \wf\ without encountering any
{\it practical\/} difficulties. Indeed, the conclusion of our analysis can
be partially summarized with the assertion that the \wf\ $\psi$ of a
subsystem represents maximal information about its configuration $X$. This
is primarily because of the \wf's statistical role, but its dynamical role
is also relevant here.  Thus it is natural, even in Bohmian mechanics, to
regard the \wf\ as the ``state'' of the system.

\rmk It has been clear, at least since von Neumann\recite{vN}, that for
all practical purposes the \qf, regarded in strictly operational terms, is
consistent. However, it has not, at least for many (e.g., Einstein), been
clear that the ``full'' \qt, regarded as including the assertion of
``completeness'' based upon Heisenberg's uncertainty principle---which has
itself traditionally been regarded as arising from the apparent
impossibility of certain measurements described in more or less {\it
classical\/} terms---is also consistent. (See\recite{Scully} for a
recent expression of related concerns.) If nothing else, \qb\ establishes
and makes clear this consistency---even including absolute uncertainty.

Indeed, as is well known, Einstein tried for many years to devise thought
experiments in which the limitations expressed by the uncertainty principle
could be evaded. The reason Einstein persisted in this endeavor is
presumably connected with the fact that the arguments presented by
Heisenberg and Bohr against such a possibility were, to say the least, not
entirely convincing, relying, as they did, on a peculiar, nearly
contradictory, combination of quantum and classical ``reasoning.'' In this
regard, recall that in order to rescue (a version of) the uncertainty
principle from one of Einstein's final onslaughts (see\recite{BE}), Bohr
felt compelled to exploit certain effects arising from Einstein's general
theory of relativity\recite{BE}.

However, from the perspective of a Bohmian universe the uncertainty
principle is sharp and clear. In particular, from such a perspective it
makes no sense to try to devise {\it thought\/} experiments by means of
which the uncertainty principle can be evaded, since this principle is a
mathematical consequence of \qb\ itself.  One could, of course, imagine a
universe governed by different laws, in which the uncertainty principle,
and a great deal else, {\it would\/} be violated, but there can be no
universe governed by \qb---and in \qe--- which fails to embody absolute
uncertainty and the uncertainty principle which it entails.

\rmk The notion of effective \wf\ developed in Section 5 should perhaps be
compared with a related notion of Bohm, namely, the ``active'' piece of the
\wf\ ~[\rcite{Bohm87},\rcite{Bohmnewbook}] (see also
Bohm\recite{Bohm1}): If $\Psi$ is of the form
\eq{Psum} with the supports of $\Psi^{(1)}$ and $\Psi^{(2)}$ ``sufficiently
disjoint,'' then $\Psi^{(i)}$ is ``active'' if the actual configuration $Q$
is in the support of $\Psi^{(i)}$. (See \eq{vrestr} and the surrounding
discussion.) When this active \wf\ appropriately factorizes---see
\eq{prod}---the (active) \wf\ of a subsystem could be defined in terms of the
obvious factor.

This notion of subsystem \wf\ will agree with ours if, as is likely to be
the case, the active and inactive pieces have suitably disjoint
$y$-supports, and it will otherwise disagree. (In this regard see also Point
20.) For example, if
$$
\Psi^{(i)}(x,y)=\psi^{(i)}(x)\Phi(y)
\eqtag{A.1}$$
with $\psi^{(1)}$ and $\psi^{(2)}$ suitably disjoint (e.g., because the
$x$-system is macroscopic and ...) then the ``active'' \wf\ of the $x$-system
is the appropriate $\psi^{(i)}$, while using our notion the $x$-system has
effective \wf\ $\psi^{(1)}+\psi^{(2)}$. Note, in particular, that with our
notion the effective \wf\ of the universe is the universal \wf\ $\Psi$, not
the active piece of $\Psi$.

Our notion of effective \wf---and not the notion based upon the active
piece---has a distinctly epistemological aspect: While for both choices we
have that ``$\born$'', the latter will be the conditional distribution given
the configuration of the environment only if $\psi$ agrees with our
effective (or conditional) \wf.  Moreover, whenever we can be said to {\it
``know\/} that the $x$-system has \wf\ $\psi$,'' then the $x$-system indeed has
effective \wf\ $\psi$ in our sense.

Note that while both of these choices are somewhat vague, in that they
appeal to the notion of the ``macroscopic''---or to some such notion---our
effective \wf, when it exists, is, as we have seen, completely unambiguous.
Moreover, as we have also seen, with our notion reference to something like
the macroscopic is not critical. Removing such a reference---as we did in
defining the notion of the conditional \wf---leads to a precise formulation
which remains entirely adequate (in fact, perfect) for our purposes. But
for the choice based on the active piece, removing such a reference would
lead to utter vagueness.

There is, of course, no real physics contingent upon a particular choice of
(notion of) ``effective \wf''; rather this choice is simply a matter of
convenience of expression, of how we talk most efficiently about the
physics. But such considerations can be quite important!

\rmk Sometimes it is helpful to try to imagine how things appear to God.
This is of course audacious, but, in fact, the very activity of a
physicist, his attempting to find the deepest laws of nature, is nothing if
not audacious.  Indeed, one might even argue that the defining activity of
the physicist is the search for the divine perspective.

Be that as it may, to create a universe God must first decide upon the
ontology---on what there is---and then on the dynamical laws---on how what
is behaves. But this alone would not be sufficient.  What is missing is a
particular realization, out of all possible solutions, of the
dynamics---the one corresponding to the actual universe. In other words, at
least for a deterministic theory, what is further required is a choice of
initial conditions. And unless there is somehow a natural special choice,
the simplest possibility would appear to be a completely random initial
condition, with an appropriate natural measure for the description of this
randomness (whatever this might mean, even given the measure). The notion
of typicality so defined would, in a sense, be an essential ingredient of the
theory governing this hypothetical universe.

For \qb, {\it with somehow given initial \wf\ $\Psi_0$\/}, this measure of
typicality is given by the \qe\ distribution ${|\Psi_0|}^2$. Moreover,
the dynamics itself is also generated by $\Psi_0$. It seems most fitting that
God should design the universe in so efficient a manner, that a single
object, the \wf\ $\Psi_0$, should generate all the necessary
(extra-ontological) ingredients.

\rmk Regarding the question of universal initial conditions, we should
perhaps contrast the issue of the initial configuration with that of the
initial \wf. Insofar as the latter is a nonequilibrium \wf, the initial
\wf\ must correspond to low entropy---it must be very atypical, i.e., of a
highly improbable character. As has been much emphasized by R. Penrose
\recite{enm}, in order to understand our nonequilibriuim world we must face the
problem of why God should have chosen such improbable initial conditions as
demanded by nonequilibrium. On the other hand, for the universal initial
configuration---in \qe---we of course have no such problem. On the
contrary, quantum randomness itself, including even absolute uncertainty,
arising as it does from \qe, in effect requires no explanation. (Concerning
the choice of initial universal \wf, see also Point 13.)

\rmk Naive agreement with the \qf\ {\it demands\/} the existence of a small set
of bad initial configurations, corresponding to outcomes which are very
unlikely but {\it not\/} impossible. It is thus hard to see how our results
could be improved upon or significantly strengthened.

More generally, for any theory with probablistic content, particularly one
describing a relativistic universe, we arrive at a similar conclusion: Once
we recognize that there is but one world (of relevance to us), only one
actual space-time history, we must also recognize that the ultimate meaning
of probability, insofar as it employed in the formulation of the
predictions of the theory, must be in terms of a specification of
typicality---one such that theoretically predicted empirical distributions
are typical.  When all is said and done, the physical import of the theory
must arise from its provision of such a notion of typical space-time
histories (at the very least of ``macroscopic'' events), presumably
specified via a probability distribution on the set of all (kinematically)
possible histories. And given a theory, i.e., such a probability
distribution, describing a large but finite universe, atypical space-time
histories, with empirical distributions disagreeing with the theoretical
predictions, are, though extremely unlikely, not impossible.

\rmk It is quite likely that the fiber $\Q_t^Y\equiv\bigl\{Q\in\Q
\bigm|Y_t=Y \bigr\}$ of $\Q$ for which $Y_t=Y$, discussed in Section 7, is
extremely small, owing to the expansive and dispersive effects of the Laplacian
$\boldsymbol\Delta$ in \Sc's equation. If so, it follows that
any regular (continuous) $\Psi_0$ (or ${|\Psi_0|}^2$) should be
approximately constant on $\Q_t^Y$ (as on any sufficiently small set of
initial conditions). This would imply that $\P_t^Y$, the conditional
measure given $\Q_t^Y$, should be approximately the same as the uniform
distribution---Lebesgue measure---on $\Q_t^Y$, so that typicality defined
in terms of \qe\ agrees with typicality in terms of Lebesgue measure.

Now, as we have already indicated in Section 4, under more careful scrutiny
this argument does not sustain its appearance of relevance. However, it may
nonetheless have some heuristic value.

\rmk We wish to emphasize that a byproduct of our analysis, quite aside
from the relevance of this analysis to the interpretation of \qt, is the
clarification and illumination of the meaning and role of probability in a
deterministic (or even nondeterministic) universe. Moreover, our analysis
of statistical tests in Section 7---the very triviality of this analyis,
see equations \eq{test} and \eq{atest}---sharply underlines the centrality
of typicality in the elucidation of the concept of probability.

\rmk We should mention some examples of nonequilibrium (initial) universal
\wf s: 

(1) Suppose that physical space is finite, say the $3$-torus ${\Bbb T}^3$
rather than $\R^3$, and suppose, say, that the potential energy $V=0$. Let
$\Psi_0\pvec \bold q,N=1$ if all $\bold q_i\in B$, where $B\subset {\Bbb
T}^3$ is a ``small'' region in physical space, and be otherwise $0$. Then
$\Psi_0$ is a nonequilibrium \wf, since an equilibrium \wf\ should be
``spread out'' over ${\Bbb T}^3$. Moreover the initial \qe\ distribution on
configurations is uniform over configurations of $N$ particles in $B$.

More generally, any well localized $\Psi_0$ is a nonequilibrium \wf. And if
physical space is $\R^3$, any localized or square-integrable \wf\ is a
nonequilibrium \wf.

(2) For a nonequilibrium \wf\ of a rather different character, consider
the following: Take ${\Bbb T}^3$ again for physical space, but instead of
considering free particles, suppose that $V$ arises from Coulomb
interactions, with half of the particles having charge $+e$ and half $-e$.
Now suppose that $\Psi_0$ is constant, $\Psi_0=1$ on ${\Bbb T}^3$. (Thus,
\qe\ now initially corresponds to a uniform distribution on
configurations.) That this $\Psi_0$, though ``spread out,'' is nevertheless
a nonequilibrium \wf\ can be seen in various ways. Dynamically, the \Sc\
evolution should presumably lead to the formation of ``atoms,'' of suitable
pairing in the (support properties of the) \wf. Entropically, $\Psi_0$ is very
special. An equilibrium ensemble of initial \wf s is determined by the
values of the infinite set of constants of the motion given by the absolute
squares of the amplitudes with respect to a basis of energy eigenfunctions.
Wave functions in this ensemble are then specified by the phases of these
amplitudes. A random choice of phases leads to an equilibrium \wf, which
should reflect the existence of ``atoms.'' On the other hand, the \wf\
$\Psi_0=1$ corresponds to a particular, very special choice of phases, so
that ``atoms cancel out.''

Note also that this example is relevant to the Penrose problem mentioned in
Point 9. What choice of initial \wf\ could be simpler---and thus in a
sense more natural---than the one which is everywhere constant? And, again,
while it might at first glance seem that this choice corresponds to
equilibrium, the attractive (in both senses) effects of the Coulomb
interaction presumably imply that this is not so!

From a classical perspective the situation is similar: The initial state in
which the particles are uniformly distributed in space with velocities all $0$
(or with independent Maxwellian velocities) is a nonequilibrium state. In
fact, an infinite amount of entropy can be extracted from suitable
clustering of the particles, arising from the great volume in momentum
space liberated when pairs of oppositely charged particles get close. (Of
course, for Newtonian gravitation---as well as for general
relativity---this tendency to cluster is, in a sense, far stronger still.)

\rmk To account for (the) most (familiar) applications of the \qf\ one
rarely needs to apply (the conclusions of) our \qe\ analysis to systems of
the form \eq{sysapp}: Randomness in the result of even a quantum
measurement usually arises solely from randomness in the system, randomness
in the apparatus making essentially no contribution. This is because most
real-world measurements are of the scattering-detection type---and a
particle (or atom ...) will be detected more or less where it's at. Think,
for example, of a two-slit-type experiment, or of the purpose of a cloud
chamber, or of a Stern-Gerlach measurement of spin.

\rmk When all is said and done, what does the incorporation of actual
configurations buy us? A great deal! It accounts for:\roster
\item randomness
\item absolute uncertainty
\item the meaning of the  \wf\ of a (sub)system
\item collapse of the wave packet
\item coherent---indeed, familiar---(macroscopic) reality
\endroster
Moreover, it makes possible an appreciation of the basic significance of
the universal \wf\ $\Psi$, as an embodiment of {\it law\/}, which cannot be
clearly discerned without a coherent ontology to be governed by some law.

\rmk Recall that in principle the \wf\ $\psi$ of a (sub)system could depend
upon the universal \wf\ $\Psi$ and on the choice of system
$\sigma=(\pi,T)$, as well as on the configuration $Y$ of the environment of
this system. In practice, however, in situations in which we in fact know
what $\psi$ is, it must be given by a function of $Y$ alone, not depending
upon $\sigma$, nor even on $\Psi$ (for ``reasonable'' nonequilibrium
$\Psi$). After all, what else, beyond $Y$, do we have at our disposal to
take into account when we conclude that a particular system has \wf\
$\psi$? In particular, $\Psi$ is unknown, apart from what we can
conclude about it on the basis of $Y$ (and perhaps some a priori assumptions
about reasonable initial $\Psi_0$'s. But even if $\Psi_0$ were known
precisely, this information would be of little use here, since solving
\Sc's equation to obtain $\Psi$ would be out of the question!)

Thus, whatever we can in practice conclude about $\psi$ must be based upon a
{\it universal\/} function---of $Y$. It would be worthwhile to explore and
elucidate the details of this function, analyzing the rules we follow in
obtaining knowledge and trying to understand the validity of these rules.
However, such considerations are not directly relevant to our purposes in this
paper, where our goal has been primarily to establish sharp {\it
limitations\/} on the possibility of knowledge rather than to analyze what
renders it at all possible. We have argued that the latter problem is
perhaps far more difficult than the former, and, indeed, that this is not
terribly astonishing.

\rmk In view of the similarity between \qb\ and stochastic mechanics [\rcite{Nelson Phys Rev},\rcite{Nelsonbm},\rcite{Quantum Fluct}], for
which similarity see~[\rcite{SGJSP},\rcite{Ascona}], all of our arguments
and results can be transferred to stochastic mechanics without significant
modification. More important, the motivation for stochastic mechanics is
the rather plausible suggestion that quantum randomness might originate
from the merging of classical dynamics with intrinsic randomness, as
described by a diffusion process, and with ``noise'' determined by $\h$.
Insofar as our results demonstrate how quantum randomness naturally emerges
without recourse to any such ``noise,'' they rather drastically erode the
evidential basis of stochastic mechanics.

\rmk The analyis of \qb\ presented here is relevant to the problem of the
interpretation and application of \qt\ in cosmology, specifically, to the
problem of the significance  of $\born$ on the cosmological level---where there
is nothing outside of the system to perform the measurements from which
$\born$ derives its very meaning in orthodox \qt.

\rmk Our random system analysis illuminates the flexibility of \qb:
It illustrates how joint probabilities as predicted by the \qf, even for
configurations, may arise from measurement and bear little resemblance to
the probabilities for unmeasured quantities. And our analysis highlights
the mathematical features which make this possible. This flexibility could
be quite important for achieving an understanding of the relativistic
domain, where it may happen that \qe\ prevails only on special space-time
surfaces (see\recite{Ascona}). Our (random system) multitime analysis
illustrates how this need entail no genuine obstacle to obtaining the
\qf.  (Our argument here of course involved the natural hypersurfaces given by
$\{t=\text{const.}\}$, but the only feature of these surfaces critical to
our analysis was the validity of \qe, or, more precisely, of the
fundamental conditional probability formula \eq{cp}.)

\rmk A notion intermediate between that of the effective \wf\ and that of
the conditional \wf\ of a subsystem, a {\it more-general-effective \wf\/}
which like the effective \wf\ is ``stable,'' may be obtained by replacing,
in the definition \eq{ewfa}--\eq{ewfb} of effective \wf, the reference to
macroscopically disjoint $y$-supports by ``sufficiently disjoint''
$y$-supports. This notion of more-general-effective
\wf\ is, of course, rather vague. But we wish to emphasize that the
$y$-supports of $\Phi$ and $\Psi^{\perp}$ may well be sufficiently disjoint
to render negligible the (effects of) future interference between the terms
of \eq{ewfa}---so that if \eq{ewfb} is satisfied, $\psi$ will indeed fully
function dynamically as the \wf\ of the $x$-system---without their having to
be actually macroscopically disjoint.

In fact, owing to the interactions---expressed in \Sc's equation---among
the many degrees of freedom, the amount of y-disjointness in the supports
of $\Phi$ and $\Psi^{\perp}$ will typically tend to increase dramatically
as time goes on, with, as in a chain reaction, more and more degrees of
freedom participating in this disjointness
(see~[\rcite{Bohm1},\rcite{Leggett},\rcite{Zurek},\rcite{JZ}]; see
also\recite{Bohmqt})).  When the effects of this dissipation or
``decoherence'' are taken into account, one finds that a small amount of
y-disjointness will often tend quickly to become ``sufficient,'' indeed
becoming ``much more sufficient'' as time goes on, and very often indeed
becoming macroscopic. Moreover, if ever we are in the position of knowing
that a system has more-general-effective \wf\ $\psi$, then $\psi$ must be
its effective \wf, since our knowledge must be based on or grounded in
macroscopic distinctions (if only in the eye or brain).

Concerning dissipation, we wish also to emphasize that in practice the
problem is not how to arrange for it to occur but how to keep it under
control, so that superpositions of (sub)system \wf s retain their coherence
and thus may interfere.

\rmk If we relax the condition \eq{psii}, requiring that $\psi_{\sigma_i}$
be nonrandom, and stipulate instead merely that 
$$
\psi_{\sigma_i}\in\Cal F\pvec Z,{i-1},
\eqtag{A.2}$$
we find that $\ovec Z,M$ have joint distribution given by the familiar
(Wigner) formula\recite{Wignermeas} (see also\recite{vN} and\recite{ABL}).

\rmk We wish to compare (what we take to be the lessons of) \qb\ with the
approach of Gell-Mann and Hartle (GMH)~[\rcite{GMH1},\rcite{GMH2}].
Unhappy about the irreducible reference to the observer in the orthodox
formulation of \qt, particularly insofar as cosmology is concerned, they
propose a program to extract from the \qf\ a ``quasiclassical domain of
familiar experience,'' which, if we understand them correctly, defines for
them the basic ontology of \qt. This they propose to do by regarding the
Wigner formula (referred to in Points 4 and 21), for the joint
probabilities of the results of a sequence of measurements of quantum
observables, as describing the probabilities of objective, i.e.,
not-necessarily-measured, events---what they call alternative histories. Of
course, owing to interference effects one quickly gets into trouble here
unless one restricts {\it this\/} use of the Wigner formula to what they
call alternative (approximately) {\it decohering\/} histories, for which
the Wigner formula can indeed be regarded as defining (approximate)
probabilities, which are additive under coarse-graining. Thus far GMH in
essence reproduce the work of Griffiths\recite{G} and Omnes\recite{O}. But,
as GMH further note, the condition of (approximate) decoherence by itself
allows for far too many possibilities. They thus introduce additional
conditions, such as ``fullness'' and ``maximality,'' as well as propose
certain (as yet tentative) measures of ``classicity'' to define an
optimization procedure they hope will yield a more or less unique
quasiclassical domain. (They also consider the possibility that there may
be many quasiclassical domains, each of which would presumably define a
different physical theory.)

As in our analysis of \qb, universal initial conditions---for GMH the
initial universal \wf (or density matrix)---play a critical role. And just
as in \qb, the \wf\ does not provide a complete description of the
universe, but rather attains physical significance from the role it plays
in generating the behavior of something else, something {\it physically\/}
primitive---for GMH the quasiclassical domain. 

Insofar as nonrelativistic \qt\ is concerned, a significant difference
between \qb\ and the proposal of GMH is that the latter defines a research
program while the former is an already existing, and sharply formulated,
physical theory. And as far as relativistic \qt\ is concerned, we believe
that, appearances to the contrary notwithstanding, the lesson of \qb\ is
one of flexibility (see also Point 19) while the approach of GMH is rigid.
In saying this we have in mind, on the one hand, that GMH insist (1) that
the possible ontologies be limited by the usual quantum description, i.e.,
correspond to a suitable (possibly time-dependent) choice of self-adjoint
operators on Hilbert space; and (2) that this ontology be constrained
further by the
\qf, demanding that its evolution be governed by the Wigner formula---so
that for them, but not for \qb, the consideration of decoherence indeed becomes
essential, bound up with questions of ontology.

On the other hand, one lesson of \qb\ is that ontology need not be so
constrained. While the \qf\ must---and for \qb\ does---emerge in
measurement-type situations, the behavior of the basic variables,
describing the fundamental ontology, outside of these situations need bear
no resemblance to anything suggested by the \qf. (Recall, in fact, that it
quite frequently happens that simple, symmetric laws on a deeper level of
description lead to a less symmetric phenomenological description on a
higher level.) Indeed, these basic variables, whether they describe
positions, or field configurations, or what have you, need not even
correspond to self-adjoint operators. That they rather trivially do in \qb\
is, in part, merely an artifact of the equivariant measure's being a
strictly local functional of the \wf, which was in no way crucial to our
analysis.

In particular, while dissipation or decoherence are relevant both to \qb\
and to GMH, for GMH they are crucial to the {\it formulation\/} of the
theory, to the specification of an {\it ontology\/}, while for \qb\ they
are relevant only on the level of {\it phenomenology\/}. And insofar as the 
formation of new theories is concerned, the lesson of \qb\ is to look for 
fundamental microscopic laws appropriate to the (or a) natural choice of
ontology, rather than to let the ontology itself be dictated by some law, let
alone by what is usually regarded as a macroscopic measurement formalism.

It is perhaps worth considering briefly the two-slit experiment. In \qb\
the electron, indeed, goes through one or the other of the two slits, the
interference pattern arising because the arrival of the electron at the
``photographic'' plate reflects the interference profile of the \wf\
governing the motion of the electron. In particular, and this is what we
wish to emphasize here, in \qb\ a spot appears somewhere on the plate
because the electron arrives there; while for GMH ``the electron arrives
somewhere'' because the spot appears there.

\rmk There is one situation where we may, in fact, know more about
configurations than what is conveyed by the \qeh\ $\born$: when we
ourselves are part of the system! See, for example, the paradox of Wigner's
friend\recite{Wignerconsc}. In thinking about this situation it is
important to note well that, while it may be merely a matter of convention
whether or not we choose to include say ourselves in the subsystem of
interest, the \wf\ to which the \qeh\ refers---that of the
subsystem---depends crucially on this choice.

\rmk We have shown, in part here and in part in\recite{op paper}, how the  \qf\
emerges within a Bohmian universe in \qe. Thus, evidence for the \qf\ is
evidence for \qe---global \qe. This should be contrasted with the
thermodynamic situation, in which the evidence points towards pockets of
\te\ within global thermodynamic nonequilibrium.

The reader may wish to explore quantum nonequilibrium.  What sort of
behavior would emerge in a universe which is initially in quantum
nonequilibrium? What phenomenological formalism or laws would govern such
behavior? We happen to have no idea! We know only that such a world is not
our world!  Or do we?
\newline\newline

\resetall\heading{Acknowledgements}\endheading

We are very grateful to Jean Bricmont, Gregory Eyink, Rebecca Goldstein,
and Eugene Speer for their aid and encouragement, and for many valuable
suggestions.  We also wish to thank Karin Berndl, Martin Daumer, Pedro
Garrido, Doug Hemmick, Martin Kruskal, Antti Kupiainen, Reinhard Lang, Joel
Lebowitz, Tim Maudlin, Giuseppe Olivieri, Herbert Spohn, and Hector
Sussmann for their assistance. We would like to acknowledge the hospitality
of the Institut des Hautes \'Etudes Scientifique, Bures-sur-Yvette, where
the basic idea for this work was conceived, and of the Fakult\"at f\"ur
Mathematik, Universit\"at M\"unchen, where much of the work was done.
Finally, we would like to thank a referee for a very careful reading; his
comments have prompted what we believe is an improved version of this
paper. The research of D.D. was supported in part by DFG, that of S.G. by
NSF grants DMS--8903047 and DMS--9105661, and that of N.Z. by DFG and
INFN.

\end
\Refs
\openup1\jot
\ref\retag{ABL} \by Y. Aharonov, P. G. Bergmann, and J. L. Lebowitz \paper
Time symmetry in the quantum process of measurement \jour Physical Review
B \vol 134 \pages 1410--1416 \yr1964 \finalinfo reprinted in\recite{WZ}
\endref

\ref\retag{Bellrmp} \by J. S. Bell \paper On the problem of hidden
variables in quantum mechanics \jour Reviews of Modern Physics \vol 38 \yr
1966 \pages 447--452 \finalinfo reprinted in\recite{WZ} and
in\recite{Bellbook}
\endref

\ref\retag{Bellepr} \by J. S. Bell \paper On the Einstein Podolsky Rosen
paradox \jour Physics \vol 1 \yr 1964 \pages 195--200 \finalinfo reprinted
in\recite{WZ} and in\recite{Bellbook}
\endref

\ref\retag{Bellepw} \by J. S. Bell \paper The measurement theory of Everett
and de Broglie's pilot wave \inbook Quantum Mechanics, Determinism,
Causality, and Particles \eds L. de Broglie and M. Flato \publ
Dordrecht-Holland, D. Reidel \publaddr Boston \yr 1976 \pages 11--17 \finalinfo
reprinted in\recite{Bellbook}
\endref

\ref\retag{Belldm} \by J. S. Bell \paper De Broglie-Bohm, delayed-choice
double-slit experiment, and density matrix \jour International Journal of
Quantum Chemistry: A Symposium \vol 14 \yr1980 \pages 155--159 \finalinfo
reprinted in\recite{Bellbook}
\endref

\ref\retag{Bertlmann's socks} \by J. S. Bell \paper Bertlmann's socks and
the nature of reality \jour Journal de Physique, C2\vol 42 \yr1981\pages
41--61 \finalinfo reprinted in\recite{Bellbook}
\endref

\ref\retag{Bellqmcosm} \by J. S. Bell \paper Quantum mechanics for
cosmologists \inbook Quantum Gravity 2 \eds C. Isham, R. Penrose, and D.
Sciama \publ Oxford University Press \publaddr New York \yr 1981
\pages 611--637 \finalinfo reprinted in\recite{Bellbook}
\endref

\ref\retag{Bellpw} \by J. S. Bell \paper On the impossible pilot wave \jour
Foundations of Physics \vol 12 \yr1982 \pages 989--999 \finalinfo reprinted
in\recite{Bellbook} 
\endref

\ref\retag{Bellj} \by J. S. Bell \paper Are there quantum jumps? \inbook
\Sc. Centenary celebration of a polymath \ed C. W. Kilmister \publ
Cambridge University Press \publaddr Cambridge \yr1987 \finalinfo reprinted
in\recite{Bellbook}
\endref

\ref\retag{Bellbook} \by J. S. Bell \book Speakable and unspeakable in
quantum mechanics \publ Cambridge University Press \publaddr Cambridge \yr
1987
\endref

\ref\retag{AM} \by J. S. Bell \paper Against ``measurement'' \jour Physics
World \vol 3 \yr 1990 \pages 33--40 \finalinfo also in\nofrills\moreref
\book Sixty-two Years of Uncertainty: Historical, Philosophical, and
Physical Inquiries into the Foundations of Quantum Mechanics \ed \ A. I.
Miller \publ Plenum Press \publaddr New York, 1990, pp. 17--31
\endref 

\ref\retag{Bohmqt} \by D. Bohm \book Quantum Theory \publ Prentice-Hall
\publaddr Englewood Cliffs, N.J. \yr1951
\endref

\ref\retag{Bohm1} \by D. Bohm \paper A suggested interpretation of the
quantum theory in terms of ``hidden variables'': Part I \jour Physical Review
\vol 85 \pages 166--179 \yr 1952 \finalinfo reprinted in \recite{WZ}
\endref

\ref\retag{Bohm2} \by D. Bohm \paper A suggested interpretation of the
quantum theory in terms of ``hidden variables'': Part II \jour Physical Review
\vol 85 \pages 180--193 \yr 1952 \finalinfo reprinted in \recite{WZ}
\endref

\ref\retag{Bohm3} \by D. Bohm \paper Proof that probability density
approaches $\psisq$ in causal interpretation of \qt \jour Physical Review
\vol 89 \pages 458--466 \yr1953 
\endref

\ref\retag{Bohm75} \by D. Bohm and B. J. Hiley \paper On the intuitive
understanding of non-locality as implied by \qt \jour Foundations of
Physics \vol 5 \pages 93--109\yr1975
\endref

\ref\retag{Bohm imp order} \by D. Bohm \book Wholeness and the Implicate
Order \publ Routledge \& Kegan Paul \publaddr London \yr1980
\endref 

\ref\retag{Bohm84} \by D. Bohm and B. J. Hiley \paper Measurement
understood through the quantum potential approach \jour Foundations of
Physics \vol 14 \yr1984 \pages 255--274
\endref

\ref\retag{Bohm87} \by D. Bohm and B. J. Hiley \paper An ontological basis
for the quantum theory I: Non-relativistic particle systems \jour Physics
Reports \vol 144 \yr1987 \pages 323--348 
\endref

\ref\retag{Bohmnewbook} \by D. Bohm and B. Hiley \book The Undivided
Universe: an Ontological Interpretation of Quantum Theory \yr1991\finalinfo
preprint
\endref

\ref\retag{BE} \by N. Bohr \paper Discussion with Einstein on
epistemological problems in atomic physics \paperinfo in\recite{Schilpp},
pp. 199--244, reprinted in\recite{Bohr} and in \recite{WZ}
\endref

\ref\retag{Bohr} \by N. Bohr \book Atomic Physics and Human Knowledge \publ
Wiley \publaddr New York \yr1958 
\endref

\ref\retag{Born} \by M. Born \paper Quantenmechanik der Sto\ss vorg\"ange
\jour Zeitschrift f\"ur Physik \vol 38 \yr1926 \pages 803--827 \transl
English translation, {\it Quantum mechanics of collision processes,}
\book  Wave Mechanics \ed \ G. Ludwig \publ Pergamon Press
\publaddr Oxford and New York, 1968
\endref

\ref\retag{dB} \by L. de Broglie \paper A tentative theory of light quanta
\jour Philosophical Magazine \vol 47 \pages 446--458 \yr1924
\endref

\ref\retag{dB2} \by L. de Broglie \paper La nouvelle dynamique des quanta
\inbook Electrons et Photons: Rapports et Discussions du Cinqui\`eme
Conseil de Physique tenu \`a Bruxelles du 24 au 29 Octobre 1927 sous les
Auspices de l'Institut International de Physique Solvay \yr1928 \pages
105--132 \publ Gauthier-Villars \publaddr Paris
\endref

\ref\retag{MW} \by B. S. DeWitt and N. Graham, Eds. \book The Many-Worlds
Interpretation of Quantum Mechanics \publ Princeton University Press
\publaddr Princeton, N.J. \yr1973
\endref

\ref\retag{D} \by R. L.  Dobrushin \paper The description of a random field
by means of conditional probabilities and conditions of its regularity
\jour Theory of Probability and its Applications \vol 13 \yr1968 \pages
197--224
\endref 

\ref\retag{Ascona} \by D. D\"urr, S. Goldstein, and N. Zangh\'i \paper On a
realistic theory for quantum physics \inbook Stochastic Processes, Geometry
and Physics \eds S. Albeverio, G. Casati, U. Cattaneo, D. Merlini, R.
Mortesi \publ World Scientific \publaddr Singapore \yr1990\pages 374--391
\endref

\ref\retag{op paper} \by D. D\"urr, S. Goldstein, and N. Zangh\'i \paper On
the role of operators in \qt \finalinfo in preparation
\endref

\ref\retag{der paper} \by D. D\"urr, S. Goldstein, and N. Zangh\'i \paper
The mystery of quantization \finalinfo in preparation
\endref

\ref\retag{GMH1} \by M. Gell-Mann and J. B. Hartle \paper Quantum mechanics
in the light of quantum cosmology \inbook Complexity, Entropy, and the
Physics of Information \ed W. Zurek \publ Addison-Wesley \publaddr Reading
\yr1990 \pages 425--458\moreref\book Proceedings of the
3rd International Symposium on Quantum Mechanics in the Light of New
Technology
\eds \ also in S. Kobayashi, H. Ezawa, Y. Murayama, and S. Nomura \publ
Physical Society of Japan \yr1990
\endref

\ref\retag{GMH2} \by M. Gell-Mann and J. B. Hartle \paper Alternative
decohering histories in quantum mechanics \yr1991 \finalinfo preprint
\endref

\ref\retag{GRW} \by G. C. Ghirardi, A. Rimini, and T. Weber \paper Unified
dynamics for microscopic and macroscopic systems \jour
Physical Review D\vol 34 \yr1986 \pages 470--491
\endref

\ref\retag{Gibbs} \by J. W. Gibbs \book Elementary Principles in
Statistical Mechanics \publ Yale University Press \yr1902 \moreref \publ
Dover \publaddr New York, 1960
\endref

\ref\retag{SGJSP} \by S. Goldstein \paper Stochastic mechanics and \qt
\jour Journal of Statistical Physics \vol 47 \yr 1987 \pages 645--667
\endref

\ref\retag{G} \by R. B. Griffiths \paper Consistent histories and the
interpretation of \qm \jour Journal of Statistical Physics \vol 36\pages
219--272\yr 1984
\endref

\ref\retag{Heisenberg} \by W. Heisenberg \book Physics and Philosophy \page
138 \publ Harper and Row \publaddr New York \yr1958
\endref

\ref\retag{JZ} \by E. Joos and H. D. Zeh \paper The emergence of classical
properties through interaction with the environment \jour Zeitschrift f\"ur
Physik B \vol 59 \yr1985 \pages223--243
\endref

\ref\retag{Krylov} \by N. S. Krylov \book Works on the Foundations of
Statistical Mechanics \publ Princeton University Press \publaddr Princeton,
N.J. \yr1979
\endref

\ref\retag{LL} \by L. D. Landau and E. M. Lifshitz \book Quantum Mechanics:
Non-relativistic Theory \bookinfo translated from the Russian by J. B.
Sykes and J. S. Bell \publ Pergamon Press \publaddr Oxford and New York \yr1958
\endref

\ref\retag{LR} \by O. E. Lanford, III, and D. Ruelle \jour Communications
in Mathematical Physics \vol 13 \yr1969 \pages 194--215
\endref

\ref\retag{Leggett} \by A. J. Leggett \paper Macroscopic quantum systems
and the \qt\ of measurement \jour Supplement of the Progress of Theoretical
Physics \vol 69 \yr1980 \pages 80--100
\endref

\ref\retag{LB} \by F. W. London and E. Bauer \book La Th\'eorie de
l'Observation en M\'ecanique Quantique \publ Hermann \publaddr Paris
\yr1939 \transl English translation by A. Shimony, J. A. Wheeler, W. H.
Zurek, J. McGrath, and S. McLean McGrath in\recite{WZ}
\endref

\ref\retag{Nelson Phys Rev} \by E. Nelson \paper Derivation of the \Sc\
equation from Newtonian mechanics \jour Physical Review \vol 150 \yr1966
\pages 1079--1085
\endref

\ref\retag{Nelsonbm} \by E. Nelson \book Dynamical Theories of Brownian
Motion \publ Princeton University Press \publaddr Princeton, N.J. \yr1967
\endref

\ref\retag{Quantum Fluct} \by E. Nelson \book Quantum Fluctuations \publ
Princeton University Press \publaddr Princeton, N.J. \yr 1985
\endref

\ref\retag{O} \by R. Omnes \paper Logical reformulation of \qm\ I \jour
Journal of Statistical Physics \vol 53 \pages 893--932 \yr1988 
\endref

\ref\retag{Penrose} \by R. Penrose \paper Quantum gravity and state-vector
reduction \inbook Quantum Concepts in Space and Time \eds R. Penrose
and C. J. Isham \publ Oxford University Press \publaddr Oxford
\yr1985
\finalinfo see also\recite{enm}
\endref

\ref\retag{enm} \by R. Penrose \book The Emperor's New Mind \publ Oxford
University Press \publaddr New York and Oxford \yr1989
\endref

\ref\retag{Schilpp} \by P. A. Schilpp, Ed. \book Albert Einstein,
Philosopher-Scientist \publ Library of Living Philosophers
\publaddr Evans\-ton, Ill.\yr 1949 
\endref

\ref\retag{cat paper} \by E. \Sc \paper Die gegenw\"artige Situation in der
Quantenmechanik \jour Naturwissenschaften \vol 23 \finalinfo 807--812,
823--828, 844--849 \yr1935 \transl English translation by J.
D. Trimmer \paper The present situation in quantum mechanics: a translation
of \Sc's ``cat paradox'' paper \jour Proceedings of the American
Philosophical Society \vol 124 \pages 323--338 \yr1980
\finalinfo reprinted in \recite{WZ}
\endref

\ref\retag{Sc entanglement} \by E. \Sc \paper Discussion of probability
relations between separated systems \jour Proceedings of the Cambridge
Philosophical Society \vol 31 \pages  555--563 \yr1935 \finalinfo {\bf
32\/} (1936), 446--452
\endref

\ref\retag{Schwartz} \by J. T. Schwartz \paper The pernicious influence of
mathematics on science \inbook Discrete Thoughts: Essays on Mathematics,
Science, and Philosophy \bookinfo by M. Kac, G. Rota, and J. T. Schwartz \publ
Birkhauser \publaddr Boston \yr 1986 \page 23
\endref

\ref\retag{Scully} \by M. O. Scully and H. Walther \paper Quantum optical
test of observation and complementarity in quantum mechanics \jour Physical
Review A \vol 39 \yr1989 \pages 5229--5236
\endref

\ref\retag{Stapp} \by H. P. Stapp \paper Light as foundation of being
\inbook Quantum Implications: Essays in Honor of David Bohm \eds B. J.
Hiley and F. D. Peat \publ Routledge \& Kegan Paul \publaddr London and New
York \yr1987
\endref

\ref\retag{vN} \by J. von Neumann \book Mathematische Grundlagen der
Quantenmechanik \publ Springer Verlag \publaddr Berlin\yr 1932 \transl
English translation by R. T. Beyer \book Mathematical Foundations of
Quantum Mechanics \publ Princeton University Press \publaddr Princeton,
N.J.\yr 1955 
\endref 

\ref\retag{Weinberg} \by S. Weinberg \paper Precision tests of quantum
mechanics \jour Physical Review Letters \vol 62 \pages 485---488 \yr1989
\endref

\ref\retag{WZ} \eds J. A. Wheeler and W. H. Zurek \book Quantum Theory
and Measurement \publ Princeton University Press \publaddr Princeton, N.J.
\yr 1983
\endref

\ref\retag{Wignerconsc} \by E. P. Wigner \paper Remarks on the mind-body
question \inbook The Scientist Speculates \ed I. J. Good \publ
Basic Books
\publaddr New York \yr1961 \finalinfo reprinted in \recite{Wignersymm} and in
\recite{WZ}
\endref

\ref\retag{Wignermeas} \by E. P. Wigner \paper The problem of measurement
\jour American Journal of Physics \vol 31 \pages 6--15 \yr 1963 \finalinfo
reprinted in \recite{Wignersymm} and in \recite{WZ}
\endref

\ref\retag{Wignersymm} \by E. P. Wigner \book Symmetries and Reflections
\publ Indiana University Press \publaddr Bloomington\yr1967
\endref

\ref\retag{Wignerint} \by E. P. Wigner \paper Interpretation of \qm \yr1976
\paperinfo in \recite{WZ}
\endref

\ref\retag{Wigner84} \by E. P. Wigner \paper Review of the quantum
mechanical measurement problem \inbook Quantum Optics,
Experimental Gravity and Measurement Theory \eds P. Meystre and M. O.
Scully \publ Plenum \publaddr New York \yr1983 \pages 43--63
\endref

\ref\retag{Zurek} \by W. H. Zurek \paper Environment-induced superselection
rules \jour Physical Review D \vol 26 \pages 1862--1880 \yr1982
\endref
\endRefs
